\RequirePackage{fix-cm}
\documentclass[aps,prx,reprint,superscriptaddress,longbibliography,floatfix]{revtex4-2}

\usepackage{amsmath}
\usepackage{braket}
\usepackage{graphicx}
\usepackage{capt-of}
\usepackage{xcolor}
\usepackage{microtype}
\usepackage[colorlinks=true, linkcolor=blue!50!black, citecolor=blue!50!black, urlcolor=blue!50!black]{hyperref}
\hypersetup{
    pdfauthor={Akinori Machino et al.},
    pdftitle={ZZ-Free bSWAP Gates under Strong Driving in Far-Detuned Fixed-Frequency Transmons}
}

\DeclareRobustCommand{\xGateName}[1]{\ifmmode\mathrm{#1}\else#1\fi}
\DeclareRobustCommand{\xBSWAP}{\xGateName{bSWAP}}
\DeclareRobustCommand{\xISWAP}{\xGateName{iSWAP}}
\DeclareRobustCommand{\xSWAP}{\xGateName{SWAP}}
\DeclareRobustCommand{\xCNOT}{\xGateName{CNOT}}
\DeclareRobustCommand{\xCZ}{\xGateName{CZ}}
\DeclareRobustCommand{\xSQRTBSWAP}{\ensuremath{\sqrt{\xBSWAP}}}

\newcommand{\xQubitA}{\ensuremath{\mathrm{Q}_{\mathrm{A}}}}
\newcommand{\xQubitB}{\ensuremath{\mathrm{Q}_{\mathrm{B}}}}
\newcommand{\xZZ}{\ensuremath{ZZ}}
\newcommand{\xXXZ}{\ensuremath{XXZ}}
\newcommand{\xSqueezingParameter}{\ensuremath{\xi_{\mathrm{R}}^2}}
\newcommand{\xOverTwoPi}[1]{\ensuremath{#1/2\pi}}
\newcommand{\xZZText}{\textit{ZZ}}

\DeclareFontShape{OMX}{cmex}{m}{n}{<->cmex10}{}
\begin{document}

\newcommand{\xQIQB}{Center for Quantum Information and Quantum Biology, The University of Osaka, Toyonaka, Osaka, Japan}
\newcommand{\xES}{Graduate School of Engineering Science, The University of Osaka, Toyonaka, Osaka, Japan}
\newcommand{\xIST}{Graduate School of Information Science and Technology, The University of Osaka, Suita, Osaka, Japan}
\newcommand{\xQuEL}{QuEL, Inc., Hachioji, Tokyo, Japan}
\newcommand{\xRIKEN}{RIKEN Center for Quantum Computing, Wako, Saitama, Japan}
\newcommand{\xUTokyo}{Graduate School of Engineering, The University of Tokyo, Bunkyo-ku, Tokyo, Japan}

\title{\xZZText{}-Free \xBSWAP{} Gates under Strong Driving\\
in Far-Detuned Fixed-Frequency Transmons}

\author{Akinori~Machino}
\email[Contact author: ]{akinori.machino@qiqb.osaka-u.ac.jp}
\affiliation{\xES}
\affiliation{\xQIQB}
\author{Kazuhisa~Ogawa}
\affiliation{\xQIQB}
\affiliation{\xIST}
\author{Yuto~Ikegai}
\affiliation{\xQIQB}
\author{Takafumi~Miyanaga}
\affiliation{\xQIQB}
\author{Koichiro~Ban}
\affiliation{\xIST}
\author{Nilton~F.~G.~Filho}
\affiliation{\xES}
\affiliation{\xQIQB}
\author{Hidehisa~Shiomi}
\affiliation{\xQIQB}
\author{Toshi~Sumida}
\affiliation{\xQuEL}
\author{Shuhei~Tamate}
\affiliation{\xRIKEN}
\author{Yutaka~Tabuchi}
\affiliation{\xRIKEN}
\affiliation{\xQIQB}
\author{Yasunobu~Nakamura}
\affiliation{\xRIKEN}
\affiliation{\xUTokyo}
\author{Makoto~Negoro}
\email[Contact author: ]{negoro.makoto.qiqb@osaka-u.ac.jp}
\affiliation{\xES}
\affiliation{\xQIQB}
\affiliation{\xIST}

\begin{abstract}
Far-detuned frequency allocations in fixed-frequency superconducting processors mitigate spectral
crowding but constrain cross-resonance interactions at practical drive amplitudes. Here, we
demonstrate that a single strong drive induces a large $XX-YY$ interaction in the far-detuned
regime, while a weak second tone tunes the accompanying \xZZ{} phase. The interaction realizes a
$150$-ns $\xSQRTBSWAP$ gate, which can directly prepare a maximally entangled state from the ground
state. For the implementations on the same qubit pair at matched drive strengths, cross resonance
gives a shorter \xCNOT{} operation, while $\xSQRTBSWAP$ gives a $33\%$ shorter \xSWAP{} operation
with a higher IRB fidelity. The multilevel theory further predicts a regime where stronger driving
can yield a speed advantage for \xCNOT{} as well, a possibility absent in the two-level model. As an
application of the $XX-YY$ interaction, we also demonstrate two-spin squeezing beyond the standard
quantum limit through two-axis countertwisting. The interaction realized here complements cross
resonance and offers a route to efficient gate synthesis and quantum simulation of interacting spin
systems.
\end{abstract}

\maketitle

\section{Introduction}
\label{sec:introduction}

Cross-resonance (CR) gates are widely used in fixed-frequency, fixed-coupling superconducting
architectures, where driving one qubit at its neighbor's transition frequency induces an effective
$ZX$ interaction used to implement \xCNOT{}
operations~\cite{rigetti2010fully,chow2011simple,sheldon2016crosstalk,jurcevic2021demonstration}.
Far-detuned frequency allocations can mitigate spectral
crowding~\cite{inoue2026systematic,ogawa2026high}, but larger qubit-frequency separations reduce the
CR interaction strength for a given coupling and drive
amplitude~\cite{magesan2020effective,tripathi2019operation,malekakhlagh2020first}. Practical upper
limits on drive amplitude therefore motivate additional native gates that complement CR for more
efficient compilation of a broader range of quantum circuits.

In addition to CR, a range of entangling gates has been proposed and demonstrated in fixed-frequency
superconducting architectures. These approaches include conditional-phase
gates~\cite{chow2013microwave,paik2016experimental,krinner2020demonstration,ding2026frequency,shirai2026high}
and population-exchange
gates~\cite{rigetti2005protocol,poletto2012entanglement,heya2021cross,nguyen2024programmable,wei2024native}.
The \xCZ{} gate realized through a conditional-phase interaction is locally equivalent to a \xCNOT{}
operation, whereas \xISWAP{} and \xBSWAP{} belong to a different local-equivalence class under the
Cartan decomposition~\cite{zhang2003geometric}. Native exchange gates therefore provide
complementary building blocks for gate synthesis alongside CR~\cite{wei2024native,huang2023quantum}.

Stark \xISWAP{} and \xBSWAP{} gates induce coherent population exchange within the odd- and
even-parity subspaces, respectively. Both gates can be activated by microwave drives detuned from
the individual qubit transitions~\cite{wei2024native}. Stark \xBSWAP{} is particularly attractive
for far-detuned qubits because it can reach two-photon resonance without closing the large frequency
gap. However, the fixed-frequency demonstration in Ref.~\citenum{wei2024native} was performed in the
straddling regime and used a two-level approximation for its analytical description. Whether Stark
\xBSWAP{} gates could be realized under strong driving in far-detuned multilevel transmons remained
untested.

Strong off-resonant driving of multilevel transmons can generate a substantial \xZZ{} interaction
alongside the desired \xBSWAP{} exchange. The resulting conditional phase must be compensated to
realize an ideal \xBSWAP{} gate. Because the $XX-YY$ and \xZZ{} interactions commute, the \xZZ{}
phase can be controlled separately from the exchange. We exploit this property by combining the
\xBSWAP{} drive with a second tone on the other qubit to tune \xZZ{} through Stark-induced \xZZ{} by
level excursions (siZZle)~\cite{mitchell2021hardware,wei2022hamiltonian,li2026implementation}. This
approach supports both ideal \xBSWAP{} gates and a tunable combination of $XX-YY$ and \xZZ{}
interactions for quantum simulation~\cite{nguyen2024programmable}.

In this work, we develop a multilevel dressed-state theory of these interactions and experimentally
demonstrate \xBSWAP{} and $\xSQRTBSWAP$ gates in a far-detuned transmon pair. We characterize the
gates using randomized and cross-entropy benchmarking. We compare the durations and fidelities of
\xCNOT{} and \xSWAP{} implementations based on CR or $\xSQRTBSWAP$ on the same qubit pair at matched
drive strengths, showing how these native gates complement each other. We also demonstrate two-spin
squeezing beyond the standard quantum limit (SQL) using the $XX-YY$ interaction.

The paper is organized as follows. Section~\ref{sec:model} develops the multilevel theory of Stark
\xBSWAP{} and \xZZ{} control. Section~\ref{sec:experiment} presents the experimental implementation,
\xZZ{} compensation, and gate benchmarking. Section~\ref{sec:discussion} discusses complementary
native interactions and quantum-simulation applications, including two-spin squeezing.
Section~\ref{sec:conclusion} summarizes the results. The appendices provide supporting theoretical
derivations, experimental methods, and details of gate synthesis and comparison.

\section{Stark \xBSWAP{} and \xZZText{} control}
\label{sec:model}

We introduce the target \xBSWAP{} unitary and derive the $XX-YY$ interaction and resonance condition
in the dressed basis under a single drive. We then compare the two-level and multilevel models. An
additional microwave tone controls the accompanying \xZZ{} interaction.

\subsection{Target gates and generating Hamiltonian}
\label{sec:target_gate}

In the ordered two-qubit basis $\lvert 00\rangle$, $\lvert 01\rangle$, $\lvert 10\rangle$, $\lvert
11\rangle$, the ideal \xBSWAP{} and $\xSQRTBSWAP$ gates are represented by
\begin{equation}
\begin{aligned}
U_{\xBSWAP{}}&=
\begin{pmatrix}
0&0&0&\mathrm{i}\\
0&1&0&0\\
0&0&1&0\\
\mathrm{i}&0&0&0
\end{pmatrix},
\\[6pt]
U_{\xSQRTBSWAP}&=
\begin{pmatrix}
\frac{1}{\sqrt{2}}&0&0&\frac{\mathrm{i}}{\sqrt{2}}\\[3pt]
0&1&0&0\\
0&0&1&0\\[3pt]
\frac{\mathrm{i}}{\sqrt{2}}&0&0&\frac{1}{\sqrt{2}}
\end{pmatrix}.
\end{aligned}
\label{eq:bswap_target}
\end{equation}
The \xBSWAP{} gate exchanges the even-parity states $\lvert 00\rangle$ and $\lvert 11\rangle$, while
the $\xSQRTBSWAP$ gate performs a half exchange. Both leave the odd-parity states $\lvert 01\rangle$
and $\lvert 10\rangle$ unchanged. A Hamiltonian proportional to $XX-YY$ generates both \xBSWAP{} and
$\xSQRTBSWAP$~\cite{roth2017analysis}. These unitaries can be written as
\begin{equation}
\begin{aligned}
U_{\xBSWAP{}}&=\exp\!\left[\frac{\mathrm{i}\pi}{4}(XX-YY)\right],\\
U_{\xSQRTBSWAP}&=\exp\!\left[\frac{\mathrm{i}\pi}{8}(XX-YY)\right].
\end{aligned}
\label{eq:target_exchange}
\end{equation}
Here, $XX=X\otimes X$ denotes a tensor product of Pauli operators, with analogous notation for other
Pauli products. The $\xSQRTBSWAP$ gate directly prepares a maximally entangled state from the ground
state:
\begin{equation}
U_{\xSQRTBSWAP}\lvert 00\rangle=\frac{\lvert 00\rangle+\mathrm{i}\lvert 11\rangle}{\sqrt{2}}.
\label{eq:target_bell}
\end{equation}

\subsection{Single-drive \xBSWAP{} exchange}
\label{sec:multilevel_model}
\label{sec:mechanism}

\begin{figure}[tbp]
\includegraphics[width=\columnwidth]{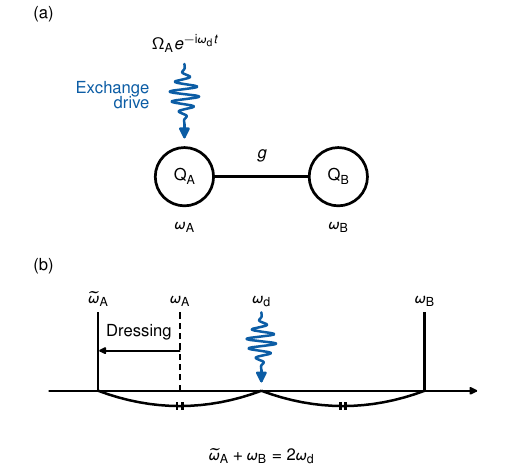}
\caption{Schematic of Stark \xBSWAP{}. (a)~Two capacitively coupled transmons with coupling $g$.
\xQubitA{} is driven with amplitude $\Omega_{\mathrm{A}}$ at frequency $\omega_{\mathrm{d}}$, while
\xQubitB{} is undriven. (b)~Dressed transition frequency $\widetilde{\omega}_{\mathrm{A}}$ of
\xQubitA{}, shifted from the bare frequency $\omega_{\mathrm{A}}$ (dashed). At the two-photon
resonance, $\widetilde{\omega}_{\mathrm{A}}$ and $\omega_{\mathrm{B}}$ are equally spaced about
$\omega_{\mathrm{d}}$, as indicated by the matching interval marks.}
\label{fig:model_schematic}
\end{figure}

Consider two capacitively coupled transmons labeled \xQubitA{} and \xQubitB{} with transition
frequencies $\omega_{\mathrm{A}}<\omega_{\mathrm{B}}$ and frequency separation
$\Delta=\omega_{\mathrm{B}}-\omega_{\mathrm{A}}$. We drive \xQubitA{} at frequency
$\omega_{\mathrm{d}}$, with drive detuning
$\delta_{\mathrm{A}}=\omega_{\mathrm{A}}-\omega_{\mathrm{d}}<0$ and real amplitude
$\Omega_{\mathrm{A}}\ge0$. Let $\lvert m\rangle$ denote the undriven eigenstates of each isolated
transmon. We define normalized transition-lowering operators
\begin{equation}
\begin{aligned}
a&=\sum_{m\ge0}\lambda_{m+1}^{\mathrm{A}}\lvert m\rangle\bra{m+1},\\
b&=\sum_{m\ge0}\lambda_{m+1}^{\mathrm{B}}\lvert m\rangle\bra{m+1},
\end{aligned}
\label{eq:transition_operators}
\end{equation}
acting on \xQubitA{} and \xQubitB{}, respectively. The real coefficients satisfy $\lambda_1^i=1$ for
$i=\mathrm{A}$ and $\mathrm{B}$. Retaining adjacent-level transitions and making the rotating-wave
approximation for the drive and coupling gives~\cite{koch2007charge,blais2021circuit}
\begin{equation}
\begin{aligned}
H_{\mathrm{rot}}&=H_{\mathrm{A}}+H_{\mathrm{B}}+V_{\mathrm{A}\mathrm{B}},\\
H_{\mathrm{A}}&=\sum_{m\ge0} E_m^{\mathrm{A}}\lvert m\rangle\bra m
+\frac{\Omega_{\mathrm{A}}}{2}(a+a^{\dagger}),\\
H_{\mathrm{B}}&=\sum_{m\ge0} E_m^{\mathrm{B}}\lvert m\rangle\bra m,\\
V_{\mathrm{A}\mathrm{B}}&=g(a^{\dagger}b+ab^{\dagger}).
\end{aligned}
\label{eq:multilevel_model}
\end{equation}
Here, we set $\hbar=1$, and $E_m^i$ denotes the undriven energy in the drive frame, with $E_0^i=0$
and $E_1^i=\delta_i$. The detuning of \xQubitB{} from the drive is
$\delta_{\mathrm{B}}=\omega_{\mathrm{B}}-\omega_{\mathrm{d}}>0$.

For analytical expressions, we use the Duffing model with anharmonicities
$\alpha_i=(E_2^i-E_1^i)-(E_1^i-E_0^i)<0$ and transition coefficients $\lambda_m^i=\sqrt m$. The
operators $a$ and $b$ then reduce to bosonic annihilation operators. For quantitative comparisons
with experiment, we retain the full cosine potential and obtain the energies $E_m^i$ and transition
coefficients $\lambda_m^i$ from the isolated transmon spectrum and charge matrix elements
(Appendix~\ref{app:cosine_model}).

Diagonalizing $H_{\mathrm{A}}$ gives dressed eigenstates $\lvert\tilde k\rangle$ and their energies
$\tilde E_k^{\mathrm{A}}$ ($k=0,1,2,\ldots$), which satisfy
\begin{equation}
H_{\mathrm{A}}\lvert \tilde k\rangle=\tilde E_k^{\mathrm{A}}\lvert \tilde k\rangle,\quad
\lvert \tilde k\rangle=\sum_{m\ge0}\langle m\vert\tilde k\rangle\,\lvert m\rangle.
\label{eq:dressed_state_expansion}
\end{equation}
The dressed states are labeled by their adiabatic connection to the undriven states. Adiabatic ramps
connect the undriven states $\lvert 00\rangle$ and $\lvert 11\rangle$ to $\lvert \tilde0,0\rangle$
and $\lvert \tilde1,1\rangle$ during the pulse and back to the undriven basis afterward
(Sec.~\ref{sec:ramp}).

To express the interaction $V_{\mathrm{A}\mathrm{B}}$ in the dressed basis, we define the matrix
elements of $a$ as
\begin{equation}
M_{kl}=\langle\tilde k\rvert a\lvert\tilde l\rangle.
\label{eq:driven_dressing}
\end{equation}
\xQubitB{} remains undriven. The transition from $\lvert \tilde0,0\rangle$ to $\lvert \tilde1,1\rangle$ raises \xQubitB{} from $\lvert 0\rangle$ to $\lvert 1\rangle$. Thus, only the $ab^{\dagger}$ term in $V_{\mathrm{A}\mathrm{B}}$ contributes to this transition. With the normalization $\lambda_1^{\mathrm{B}}=1$, the matrix element for \xBSWAP{} exchange is
\begin{equation}
\begin{aligned}
\langle\tilde1,1\rvert V_{\mathrm{A}\mathrm{B}}\lvert\tilde0,0\rangle
&=g\langle\tilde1\rvert a\lvert\tilde0\rangle
\langle1\rvert b^{\dagger}\lvert0\rangle\\
&=gM_{10}.
\end{aligned}
\label{eq:dressed_exchange_element}
\end{equation}
When the states $\lvert \tilde0,0\rangle$ and $\lvert \tilde1,1\rangle$ have equal energy, the
matrix element in Eq.~\eqref{eq:dressed_exchange_element} gives the resonant exchange Hamiltonian in
the interaction picture (Appendix~\ref{app:unified_interactions}):
\begin{equation}
\begin{aligned}
H_{XY}&=gM_{10}\bigl(
\lvert \tilde1,1\rangle\langle\tilde0,0\rvert
+\lvert \tilde0,0\rangle\langle\tilde1,1\rvert\bigr)\\
&=\frac{gM_{10}}{2}(XX-YY),
\end{aligned}
\label{eq:heff}
\end{equation}
where the Pauli operators act in the dressed computational basis. We define the exchange coefficient
as
\begin{equation}
J_{XY}=-2gM_{10}.
\label{eq:multilevel_exchange}
\end{equation}
For $J_{XY}>0$, this exchange Hamiltonian gives the $+\mathrm{i}$ phase of the exchange amplitudes
in Eq.~\eqref{eq:bswap_target}. The ideal \xBSWAP{} and $\xSQRTBSWAP$ gate durations are
$\frac{\pi}{J_{XY}}$ and $\frac{\pi}{2J_{XY}}$, respectively.

\begin{figure}[tbp]
\includegraphics[width=\columnwidth]{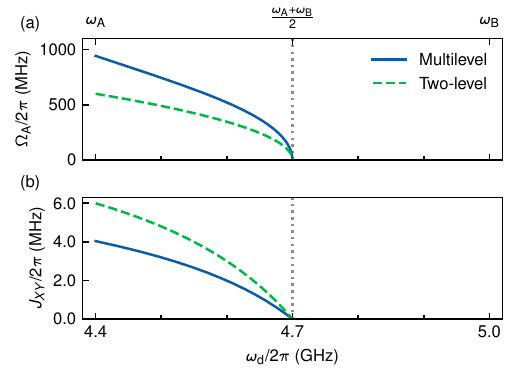}
\caption{Calculated \xBSWAP{} resonances in the two-level and multilevel transmon models. (a)~Drive
amplitude $\Omega_{\mathrm{A}}$ versus drive frequency $\omega_{\mathrm{d}}$ along the resonance.
(b)~Exchange coefficient $J_{XY}$ along each resonance curve. Dashed green and solid blue lines show
the values obtained from the two-level and multilevel models, respectively. Dotted vertical lines
mark the midpoint between the bare qubit frequencies. The model parameters are
$\xOverTwoPi{\omega_{\mathrm{A}}}=4.4\,\mathrm{GHz}$,
$\xOverTwoPi{\omega_{\mathrm{B}}}=5.0\,\mathrm{GHz}$,
$\xOverTwoPi{\alpha_{\mathrm{A}}}=-200\,\mathrm{MHz}$, and $\xOverTwoPi{g}=6\,\mathrm{MHz}$. The
full-cosine calculation uses 12 levels in \xQubitA{}.}
\label{fig:resonance_comparison}
\end{figure}

To leading order in $g$, we can obtain the \xBSWAP{} resonance condition
\begin{equation}
\bigl(\tilde E_1^{\mathrm{A}}-\tilde E_0^{\mathrm{A}}\bigr)+\delta_{\mathrm{B}}=0
\label{eq:resonance_controls}
\end{equation}
by matching the rotating-frame energies of $\lvert \tilde0,0\rangle$ and $\lvert \tilde1,1\rangle$.
By defining the dressed transition frequency of \xQubitA{} in the laboratory frame as
\begin{equation}
\widetilde{\omega}_{\mathrm{A}}\equiv\omega_{\mathrm{d}}+\tilde E_1^{\mathrm{A}}-\tilde E_0^{\mathrm{A}},
\label{eq:dressed_transition_frequency}
\end{equation}
we can rewrite the resonance condition in Eq.~\eqref{eq:resonance_controls} as
$\widetilde{\omega}_{\mathrm{A}}+\omega_{\mathrm{B}}=2\omega_{\mathrm{d}}$. This condition places
$\widetilde{\omega}_{\mathrm{A}}$ and $\omega_{\mathrm{B}}$ symmetrically about the drive frequency
$\omega_{\mathrm{d}}$, as illustrated in Fig.~\ref{fig:model_schematic}(b). Because
$\widetilde{\omega}_{\mathrm{A}}$ depends on both $\omega_{\mathrm{d}}$ and $\Omega_{\mathrm{A}}$,
this condition defines the resonance curve in Fig.~\ref{fig:resonance_comparison}(a). The resonance
condition including second-order energy shifts in $g$ is given in
Appendix~\ref{app:diagonal_shifts}.

As a concrete example, let \xQubitA{} be an ideal two-level system (see
Appendix~\ref{app:two_level_matrix} for the detailed derivation). Define the mixing angle $\theta$
by $\tan\theta=\frac{\Omega_{\mathrm{A}}}{\lvert\delta_{\mathrm{A}}\rvert}$ with
$0\le\theta<\frac{\pi}{2}$. The matrix in the ordered dressed basis $\lvert \tilde0\rangle$, $\lvert
\tilde1\rangle$ is
\begin{equation}
M=
\begin{pmatrix}
\sin\frac{\theta}{2}\cos\frac{\theta}{2}&\cos^2\frac{\theta}{2}\\[3pt]
-\sin^2\frac{\theta}{2}&-\sin\frac{\theta}{2}\cos\frac{\theta}{2}
\end{pmatrix}.
\label{eq:two_level_matrix}
\end{equation}
For the two-level model, $\tilde E_0^{\mathrm{A}}-\tilde
E_1^{\mathrm{A}}=\sqrt{\delta_{\mathrm{A}}^2+\Omega_{\mathrm{A}}^2}$. The states $\lvert
\tilde0,0\rangle$ and $\lvert \tilde1,1\rangle$ are resonant when
\begin{equation}
\sqrt{\delta_{\mathrm{A}}^2+\Omega_{\mathrm{A}}^2}=\delta_{\mathrm{B}}.
\label{eq:two_level_resonance}
\end{equation}
The element $M_{10}=-\sin^2\frac{\theta}{2}$ then gives
\begin{equation}
J_{XY}=g(1-\cos\theta)=g\left(1-\frac{\lvert\delta_{\mathrm{A}}\rvert}
{\sqrt{\delta_{\mathrm{A}}^2+\Omega_{\mathrm{A}}^2}}\right).
\label{eq:two_level_exchange_main}
\end{equation}

For a multilevel transmon, numerical diagonalization of the driven single-qubit Hamiltonian gives
the dressed energies and $M_{10}$. At each drive frequency, Eq.~\eqref{eq:resonance_controls}
determines the resonant amplitude, and $J_{XY}=-2gM_{10}$ gives the exchange coefficient.
Approximate analytical expressions within the Duffing model are given in
Appendix~\ref{app:duffing_asymptotics}.

Figure~\ref{fig:resonance_comparison} compares the two-level model with a multilevel calculation. In
the weak-drive limit, the resonance frequency approaches the midpoint between the qubit frequencies.
We instead operate in the strong-drive regime to obtain stronger exchange. The drive shifts the
transition frequency of \xQubitA{} according to Eq.~\eqref{eq:dressed_transition_frequency}, and the
two-photon resonance shifts accordingly. Along the ideal two-level resonance curve, $J_{XY}$
approaches its maximum $g$ as $\Omega_{\mathrm{A}}\to\Delta$ and
$\omega_{\mathrm{d}}\to\omega_{\mathrm{A}}$. Numerical diagonalization of the multilevel model in
Fig.~\ref{fig:resonance_comparison} instead gives $\Omega_{\mathrm{A}}\simeq1.57\Delta$ and
$J_{XY}\simeq0.67g$ at this endpoint. This difference reflects drive-induced mixing with higher
transmon levels.

\begin{table}[b]
\caption{Parameters of the \xQubitA{}--\xQubitB{} pair.}
\label{tab:device}
\centering
\setlength{\tabcolsep}{5pt}
\renewcommand{\arraystretch}{1.05}
\begin{ruledtabular}
\begin{tabular}{lcc}
 & \xQubitA{} & \xQubitB{} \\
\colrule
$\xOverTwoPi{\omega_i}$ (GHz) & $4.14753$ & $4.66328$ \\
$\xOverTwoPi{\alpha_i}$ (MHz) & $-197.0$ & $-257.2$ \\
$T_1$ (\textmu s) & $89.3$ & $88.6$ \\
$T_2^{\mathrm{echo}}$ (\textmu s) & $44.7$ & $53.7$ \\
\colrule
$\xOverTwoPi{\Delta}$ (MHz) & \multicolumn{2}{c}{$515.75$} \\
$\xOverTwoPi{g}$ (MHz) & \multicolumn{2}{c}{$5.9$} \\
$\xOverTwoPi{\zeta_0}$ (kHz) & \multicolumn{2}{c}{$-172$} \\
\end{tabular}
\end{ruledtabular}
\end{table}

\subsection{\xZZText{} control with siZZle}
\label{sec:sizzle_theory}

In multilevel transmon qubits, the $XX-YY$ interaction in Sec.~\ref{sec:multilevel_model} is
accompanied by a \xZZ{} interaction. Its coefficient $\zeta$ contains both the static \xZZ{}
coupling and the contribution induced by the drive on \xQubitA{}. The interaction Hamiltonian can be
written as
\begin{equation}
H_{\mathrm{int}}=-\frac{J_{XY}}{4}(XX-YY)+\frac{\zeta}{4}\xZZ{}.
\label{eq:exchange_with_zz}
\end{equation}
The expression for $\zeta$ is derived in Appendix~\ref{app:diagonal_shifts}.

Since the $XX-YY$ and \xZZ{} terms commute, their evolutions can be treated separately. We add a
second tone on \xQubitB{} at the \xBSWAP{} drive frequency $\omega_{\mathrm{d}}$ to control $\zeta$
through siZZle~\cite{wei2022hamiltonian}. Its amplitude $\Omega_{\mathrm{B}}$ and phase $\varphi$
relative to the drive on \xQubitA{} serve as the control parameters for the \xZZ{} interaction. With
the drive on \xQubitA{} held fixed, a weak second tone changes the coefficient according to
\begin{equation}
\zeta(\Omega_{\mathrm{B}},\varphi)\simeq\zeta(0)+\chi_{ZZ}\Omega_{\mathrm{B}}\cos\varphi,
\label{eq:sizzle_response}
\end{equation}
where $\zeta(0)$ is the value without the second tone and $\chi_{ZZ}$ is derived in
Appendix~\ref{app:sizzle_derivation}. The phase selects the sign of the change and the amplitude
sets its size. Choosing them so that $\zeta=0$ removes the \xZZ{} term and yields the ideal
\xBSWAP{} Hamiltonian in Eq.~\eqref{eq:heff}.

The same control can also be used to enhance the \xZZ{} interaction. The resulting operation
combines even-parity exchange with a conditional phase. It is the \xBSWAP{} counterpart of fSim
gates, which instead exchange populations within the odd-parity
subspace~\cite{foxen2020demonstrating}. It is locally equivalent to an \xXXZ{}-type evolution and
offers a route to Hamiltonian engineering~\cite{nguyen2024programmable}. The local equivalence,
anisotropy, and isotropic Heisenberg condition are given in Appendix~\ref{app:anisotropy}.

\begin{figure}[t]
\includegraphics[width=\columnwidth]{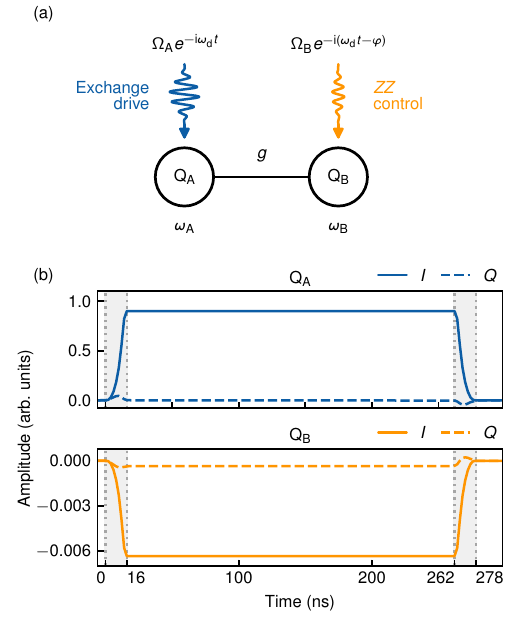}
\caption{Two-tone \xBSWAP{} control. (a)~Exchange drive on \xQubitA{} and siZZle tone on \xQubitB{}.
(b)~Pulse envelopes of the \xQubitA{}--\xQubitB{} \xBSWAP{} pulse. Solid and dashed curves show the
in-phase ($I$) and quadrature ($Q$) components, respectively, with separate vertical scales for
\xQubitA{} and \xQubitB{}. Shading and dotted boundaries mark the ramps that adiabatically connect
the undriven and dressed states.}
\label{fig:pulse_envelopes}
\end{figure}

\section{Experimental demonstration}
\label{sec:experiment}
\label{sec:calibration}
\label{sec:characterization}

We use a superconducting processor designed at RIKEN and operated at The University of Osaka. The
device comprises far-detuned, fixed-frequency transmons with fixed couplings arranged on a square
lattice. The two qubits are labeled \xQubitA{} and \xQubitB{} as in Sec.~\ref{sec:model}. Their
parameters are listed in Table~\ref{tab:device}, with transition frequencies used for qubit control.
Microwave pulses are generated by QuEL-1 SE hardware~\cite{kurimoto2026microwave} and
Qubex~\cite{machino2026qube}. Calibrated DRAG pulses implement single-qubit
gates~\cite{motzoi2009simple}. The two qubits are read out simultaneously by dispersive measurement.

The strong exchange drive is applied to \xQubitA{}, and a weak tone on \xQubitB{} at the same drive
frequency cancels the accumulated \xZZ{} phase. Figure~\ref{fig:pulse_envelopes} shows the two-tone
control and the pulse envelopes. Quasi-adiabatic ramps with a counter-diabatic quadrature connect
the undriven and dressed computational states~\cite{martinis2014fast,berry2009transitionless}. The
benchmarked \xBSWAP{} and $\xSQRTBSWAP$ gates have total durations of $278$ and $150\,\mathrm{ns}$,
respectively, including two $16$-ns ramps. Both use a drive frequency of $4.324\,\mathrm{GHz}$ and
an estimated flat-top drive amplitude of $\xOverTwoPi{\Omega_{\mathrm{A}}}\simeq419\,\mathrm{MHz}$,
inferred from the measured ac Stark shift~(Appendix~\ref{app:device_characterization}). The ramp
design and gate calibration procedures are described in Appendix~\ref{app:ramp}.
\label{sec:ramp}

\subsection{Resonant exchange under strong driving}
\label{sec:exchange}

\begin{figure}[tbp]
\includegraphics[width=\columnwidth]{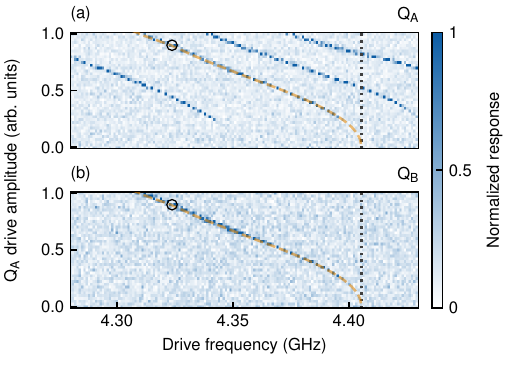}
\caption{Drive-amplitude dependence of the \xQubitA{}--\xQubitB{} resonance. The normalized readout
responses of (a)~\xQubitA{} and (b)~\xQubitB{} to a drive on \xQubitA{} are plotted. Dotted vertical
lines mark the midpoint between the undriven qubit frequencies, and open circles mark the
benchmarked drive setting ($\xOverTwoPi{\omega_{\mathrm{d}}}=4.324\,\mathrm{GHz}$,
$\xOverTwoPi{\Omega_{\mathrm{A}}}\simeq419\,\mathrm{MHz}$). Orange dashed curves show the prediction
from the multilevel transmon model. Drive amplitude is normalized to the maximum amplitude of the
control electronics.}
\label{fig:regime}
\end{figure}

The amplitude-dependent resonance predicted in Sec.~\ref{sec:mechanism} is visible directly in
spectroscopy. Figure~\ref{fig:regime} shows the responses of \xQubitA{} and \xQubitB{} to a
fixed-duration drive on \xQubitA{}. At weak drive, the common \xBSWAP{} resonance ridge lies near
the midpoint between the undriven qubit frequencies and shifts to lower frequency as the drive
amplitude increases. The marked gate setting at $4.324\,\mathrm{GHz}$ is about $81\,\mathrm{MHz}$
below the midpoint. The orange curves, calculated from the device parameters, closely follow the
measured resonance ridge. Independent Stark-Ramsey measurements calibrate the relation between
$\Omega_{\mathrm{A}}$ and the amplitude setting of the control electronics at several drive
frequencies. We interpolate this relation across drive frequency to plot the model curves
(Appendix~\ref{app:amplitude_calibration}). The full frequency window and the simultaneously
measured nearby-qubit responses are given in Appendix~\ref{app:wide_spectrum}.

\begin{figure}[tbp]
\includegraphics[width=\columnwidth]{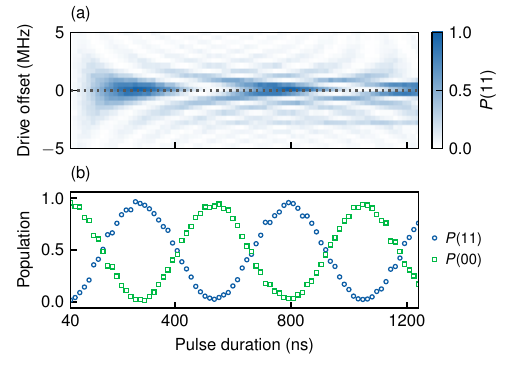}
\caption{$\lvert 00\rangle\mathord{\leftrightarrow}\lvert 11\rangle$ exchange. (a)~Chevron pattern
of the $\lvert 11\rangle$ population after preparing $\lvert 00\rangle$, as a function of the
drive-frequency offset and pulse duration. (b)~Populations of $\lvert 00\rangle$ and $\lvert
11\rangle$ at zero drive-frequency offset, indicated by the dotted horizontal line in (a).}
\label{fig:chevron}
\end{figure}

The drive activates coherent exchange between $\lvert 00\rangle$ and $\lvert 11\rangle$.
Figure~\ref{fig:chevron} shows the \xQubitA{}--\xQubitB{} populations versus drive frequency and
pulse duration at a fixed amplitude setting. The population oscillations form a chevron pattern near
$4.324\,\mathrm{GHz}$. In Fig.~\ref{fig:chevron}, readout-error mitigation is applied using
calibrated confusion matrices.

The frequency--duration scan provides the initial operating point. We then refine the pulse duration
by fitting data from repeated-pulse measurements. The local phases are removed by virtual-$Z$
updates~\cite{mckay2017efficient}. The drive frame and qubit frames rotate at different frequencies.
We compensate the phase shifts associated with the position of the gate in a circuit using the
\xBSWAP{} pulse phase and virtual-$Z$ frame updates. Appendix~\ref{sec:calibration_procedure}
describes the pulse-duration calibration and phase-correction procedures.
\label{sec:phases}

\begin{figure*}[t]
\begin{minipage}[t]{\columnwidth}
\includegraphics[width=\columnwidth]{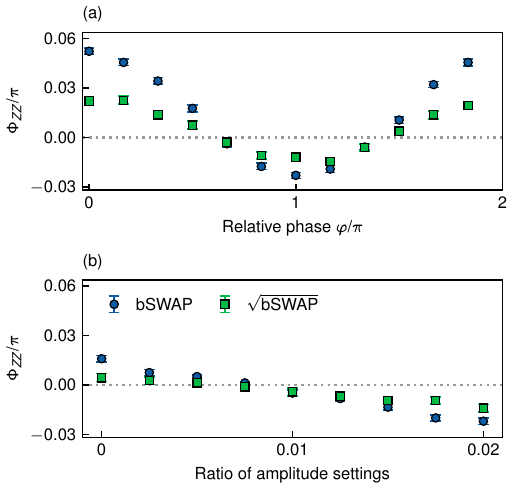}
\caption{\xZZ{}-phase control with an additional siZZle tone. Accumulated phase versus (a)~relative
phase $\varphi$ and (b)~ratio of the amplitude settings for the siZZle and \xBSWAP{} tones. Circles
denote \xBSWAP{} and squares denote $\xSQRTBSWAP$. Bars show $95\%$ statistical intervals. Dotted
lines mark zero accumulated phase.}
\label{fig:q55_zz_controls}
\end{minipage}\hfill
\begin{minipage}[t]{\columnwidth}
\includegraphics[width=\columnwidth]{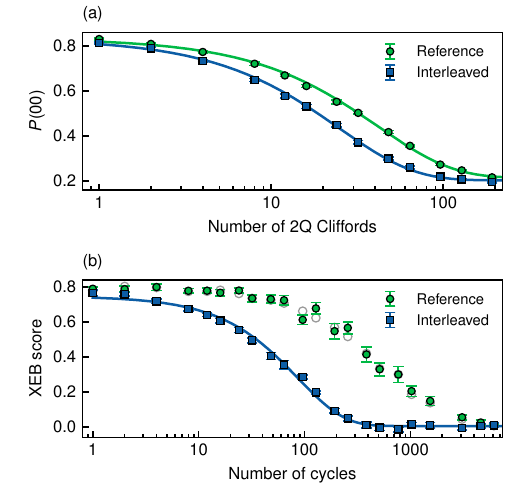}
\caption{Benchmarking of \xZZ{}-compensated gates. (a)~Interleaved randomized benchmarking of
\xBSWAP{}. (b)~Interleaved XEB of $\xSQRTBSWAP$. Filled circles and squares denote reference and
interleaved data, respectively. Bars show one standard error. Solid lines are exponential fits with
free offsets. Faint open circles in (b) show reference-model predictions.}
\label{fig:gate_benchmarks}
\label{fig:irb}
\label{fig:xeb}
\end{minipage}
\end{figure*}

\subsection{\xZZText{}-free \xBSWAP{} gates}
\label{sec:zz}
\label{sec:bench}

The \xZZ{} phase accumulated during a \xBSWAP{} pulse of duration $T$ is
\begin{equation}
\Phi_{ZZ}=\frac14\int_0^T\zeta(t)\,\mathrm{d}t,
\label{eq:phizz}
\end{equation}
corresponding to $U_{ZZ}=\exp(-\mathrm{i}\Phi_{ZZ}\xZZ{})$.
To cancel this phase, we keep the \xBSWAP{} pulse on \xQubitA{} fixed and add a siZZle tone on
\xQubitB{}. We set its relative phase $\varphi$ so that its contribution to the \xZZ{} phase is
opposite in sign to the accumulated phase, and then adjust the siZZle amplitude
$\Omega_{\mathrm{B}}$ to bring $\Phi_{ZZ}$ to zero. In independent measurements, we confirmed that
the residual \xZZ{} phases were reduced to $1.4\times10^{-3}\pi$ and $6.6\times10^{-4}\pi$ for the
\xBSWAP{} and $\xSQRTBSWAP$ gates, respectively (Appendix~\ref{app:phase_tracking}).

Figure~\ref{fig:q55_zz_controls} shows that the accumulated \xZZ{} phase can be tuned through zero
by varying the relative phase or amplitude of the siZZle tone. The phase scans use a siZZle
amplitude setting equal to $2\%$ of the \xBSWAP{} amplitude setting. In the amplitude scans,
$\varphi$ is held at its calibrated value for each gate.

\label{sec:sqrt}
\label{sec:xeb}
With the \xZZ{} phase compensated, we evaluate the Clifford \xBSWAP{} gate by interleaved randomized
benchmarking (IRB)~\cite{magesan2012efficient,corcoles2013process}. For the non-Clifford
$\xSQRTBSWAP$ gate, we instead use interleaved cross-entropy benchmarking
(XEB)~\cite{zhang2024tunable}. IRB gives a fidelity of $98.78\%$ for the $278$-ns \xBSWAP{} gate,
with a $95\%$ statistical interval of $[98.61,98.96]\%$ [Fig.~\ref{fig:gate_benchmarks}(a)]. XEB
gives a fidelity of $99.34\%$ for the $150$-ns $\xSQRTBSWAP$ gate, with a $95\%$ statistical
interval of $[99.27,99.42]\%$ [Fig.~\ref{fig:gate_benchmarks}(b)]. The benchmarking protocols and an
additional IRB measurement of a two-pulse $\xSQRTBSWAP$ block are described in
Appendix~\ref{app:benchmarks}.

\label{sec:coherence_comparison}

\begin{figure}[t]
\includegraphics[width=\columnwidth]{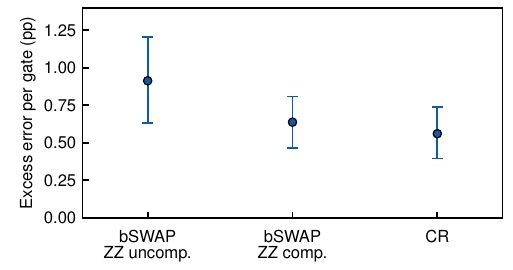}
\caption{Excess error relative to the coherence limit for \xBSWAP{} without and with \xZZ{}
compensation and echoed CR, in percentage points. Bars show $95\%$ statistical intervals with the
Table~\ref{tab:device} coherence values held fixed.}
\label{fig:coherence_gap}
\end{figure}

Figure~\ref{fig:coherence_gap} compares the excess errors relative to the coherence limit for
\xBSWAP{} with and without \xZZ{} compensation and for echoed CR. Using the $T_1$ and
$T_{2,\mathrm{echo}}$ values in Table~\ref{tab:device} with the coherence-limit expression in
Ref.~\citenum{wei2024characterizing}, we obtain reference fidelities of $99.42\%$ for the $278$-ns
\xBSWAP{} and $99.44\%$ for the $268$-ns echoed CR gate. The \xZZ{}-compensated \xBSWAP{} result is
taken from Fig.~\ref{fig:gate_benchmarks}(a). The excess errors are $0.91$ and $0.64$ percentage
points for \xBSWAP{} without and with \xZZ{} compensation, respectively, and $0.56$ percentage
points for CR.

\label{sec:bell_preparation}
We also demonstrate direct Bell-state preparation using a single $150$-ns $\xSQRTBSWAP$ pulse, which
prepares the even Bell state $(\lvert 00\rangle+\mathrm{i}\lvert 11\rangle)/\sqrt2$ from $\lvert
00\rangle$. After initial-state heralding, we reconstruct the density matrix by maximum-likelihood
estimation using the readout confusion matrix. The Bell-state fidelity is $98.74\%$, with a $95\%$
statistical interval of $[98.02,99.38]\%$. The preparation uses no siZZle tone because the \xZZ{}
interaction contributes only a global phase within the even-parity subspace.

\section{Discussion}
\label{sec:discussion}

\subsection{Complementary roles of \xBSWAP{} and CR}
\label{sec:complementary}

The benefit of a native interaction depends on the operation to be synthesized. CR supplies one
Cartan axis, while \xBSWAP{} exchange supplies two equal axes. Their ideal interaction times
therefore differ for \xCNOT{} and \xSWAP{} operations~\cite{vidal2002interaction,wei2024native}. The
\xCNOT{} construction requires two $\xSQRTBSWAP$ gates, compared with one $R_{ZX}(\frac{\pi}{2})$
gate for CR. The \xSWAP{} construction requires three native two-qubit gates in either case. These
different gate counts lead to different conditions for a speed
advantage~(Appendix~\ref{app:speed_bound}).

\suppressfloats[t]

The implementations of \xCNOT{} and \xSWAP{} on \xQubitA{}--\xQubitB{} show this target dependence
(Table~\ref{tab:compare}). The CR- and $\xSQRTBSWAP$-based implementations are compared at matched
drive amplitudes inferred from Stark shifts. CR gives a $26\%$ shorter \xCNOT{} with higher
fidelity. The $\xSQRTBSWAP$-based \xSWAP{} is $33\%$ shorter and also has higher fidelity. Selecting
the native gate for each target operation can therefore make circuit compilation more efficient.
Comparison details are given in Appendix~\ref{app:cr_comparison}.

\begin{table}[b]
\caption{\xCNOT{} and \xSWAP{} implementations. Durations $T$ include local operations and ramps.
Fidelities $\mathcal{F}$ are obtained from IRB, with one bootstrap standard error in parentheses.}
\label{tab:compare}
\begin{ruledtabular}
\begin{tabular}{lcccc}
Target & \multicolumn{2}{c}{$\xSQRTBSWAP$-based} & \multicolumn{2}{c}{CR-based} \\
 & $T$ (ns) & $\mathcal{F}$ (\%) & $T$ (ns) & $\mathcal{F}$ (\%) \\
\colrule
\xCNOT{} & $396$ & $97.69(23)$ & $292$ & $99.04(8)\phantom{0}$ \\
\xSWAP{} & $570$ & $97.00(22)$ & $852$ & $96.17(19)$ \\
\end{tabular}
\end{ruledtabular}
\end{table}

Frequency collisions provide another motivation for complementary native
interactions~\cite{inoue2026systematic,ogawa2026high}. The CR drive is tied to the target-qubit
transition, whereas the Stark \xBSWAP{} drive frequency and amplitude can be chosen together along
the resonance curve. This freedom may help avoid unwanted resonances, with a trade-off in exchange
rate and required drive amplitude.

Native \xBSWAP{} gates may also serve as building blocks for quantum error correction. Dynamic
surface-code circuits have been demonstrated using \xISWAP{} gates in place of conventional
\xCNOT{}/\xCZ{} gates~\cite{eickbusch2025demonstration}. Since \xBSWAP{} is locally equivalent to
\xISWAP{}, analogous circuits could be implemented using native \xBSWAP{} gates and local
single-qubit rotations. Compensating the accompanying \xZZ{} phase is important for realizing the
intended Clifford operation in such circuits.

\subsection{Multilevel effects on gate speed}

The analytical description of native Stark \xBSWAP{} in Ref.~\citenum{wei2024native} is based on a
two-level model. Our multilevel treatment includes drive-induced mixing with higher transmon levels.
With the same hardware limit on drive amplitude, the two-level model does not permit a
$\xSQRTBSWAP$-based construction with a shorter ideal \xCNOT{} interaction time than
CR~(Appendix~\ref{app:speed_bound}). Our multilevel model instead predicts this speed advantage at
large qubit detuning and strong drive, where $J_{XY}>2J_{ZX}$. The matrix elements determining the
\xBSWAP{} and CR coefficients respond differently to this mixing, which changes the gate-speed
comparison (Fig.~\ref{fig:speed_bound}).

Realizing this predicted regime requires control hardware capable of delivering stronger drives with
high amplitude stability. Amplitude fluctuations shift the dressed-state splitting, detuning the
two-photon resonance and changing the accumulated local phases.

\begin{figure}[t]
\includegraphics[width=\columnwidth]{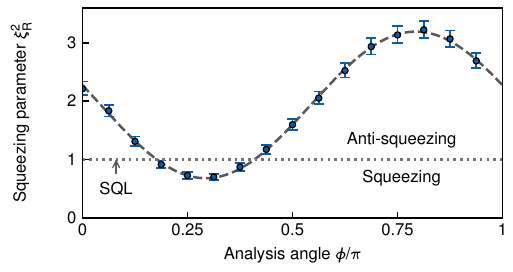}
\caption{Two-qubit spin squeezing. Data points show the Wineland spin-squeezing parameter
\xSqueezingParameter{} versus analysis angle $\phi$. This angle specifies the direction along which
transverse spin fluctuations are measured, from $x$ toward $y$ in the plane perpendicular to the
mean spin. The dashed curve is a sinusoidal fit used to determine the squeezed axis. Error bars show
$95\%$ statistical intervals. The dotted line at $\xSqueezingParameter{}=1$ marks the SQL for a
coherent state.}
\label{fig:squeezing}
\end{figure}

\subsection{Applications to quantum simulation}
\label{sec:collective_outlook}

\label{sec:squeezing}

Native access to the $XX-YY$ interaction provides a direct route to quantum simulation of
spin-squeezing dynamics. For the two-qubit collective spin
$S_\mu=(\sigma_\mu^{\mathrm{A}}+\sigma_\mu^{\mathrm{B}})/2$ with $\mu=x,y,z$, the Hamiltonian in
Eq.~\eqref{eq:heff} takes the two-axis countertwisting form~\cite{kitagawa1993squeezed},
\begin{equation}
H_{XY}=-\frac{J_{XY}}{2}(S_x^2-S_y^2).
\label{eq:collective_squeezing}
\end{equation}
Partial evolution from $\lvert00\rangle$ can reduce transverse spin fluctuations while retaining a
nonzero mean longitudinal spin.

The two-qubit measurements in Fig.~\ref{fig:squeezing} illustrate this application. We identify the
squeezed spin direction using the method of Ref.~\cite{meyer2001experimental}. An independent
measurement along this direction yields a Wineland parameter~\cite{wineland1994squeezed} of
$\xSqueezingParameter{}=0.684$, with a $95\%$ statistical interval of $[0.644,0.729]$, below the
coherent-state SQL of $\xSqueezingParameter{}=1$. This result uses initial-state heralding without
readout-error mitigation. Preparation conditions and the definition and measurement of
\xSqueezingParameter{} are given in Appendix~\ref{app:squeezing_preparation}.

Control of the accompanying \xZZ{} interaction also provides access to anisotropic spin models. On a
phase-aligned bipartite lattice, a sublattice basis change maps $XX-YY$ plus \xZZ{} to \xXXZ{}
interactions~(Appendix~\ref{app:anisotropy}).

For example, collective spin squeezing has been predicted in short-range \xXXZ{} lattices for
suitable anisotropies and initial states~\cite{perlin2020spin,roscilde2024scalable}. Coherent
averaging with isotropic exchange provides another proposed route to collective countertwisting
within a protected spin manifold~\cite{cappellaro2009quantum}.

Realizing these applications would require driving many qubit pairs simultaneously. Accessing a
wider range of interaction anisotropies would require stronger siZZle tones and joint calibration of
the siZZle and \xBSWAP{} drives.

\section{Conclusion}
\label{sec:conclusion}

We demonstrated \xZZ{}-free Stark \xBSWAP{} gates in far-detuned, fixed-frequency transmons using a
strong exchange drive and a weak tone to compensate the accompanying \xZZ{} phase. The $278$-ns
\xBSWAP{} had a fidelity of $98.78\%$ from IRB, and the $150$-ns $\xSQRTBSWAP$ had a fidelity of
$99.34\%$ from interleaved XEB. A single $\xSQRTBSWAP$ pulse directly prepared a Bell state from the
ground state. The $XX-YY$ interaction also generated two-spin squeezing below the SQL.

The \xCNOT{} and \xSWAP{} comparisons demonstrated the complementary roles of CR and $\xSQRTBSWAP$.
CR gave a shorter \xCNOT{} with higher fidelity, while the $\xSQRTBSWAP$-based \xSWAP{} was $33\%$
shorter and also had higher fidelity. Selecting the native gate for each target operation can
therefore make circuit compilation more efficient. Control of $XX-YY$ and \xZZ{} interactions also
offers a route to quantum simulation of interacting spin systems.

\begin{acknowledgments}
This work was supported in part by JST COI-NEXT (Grant No.\ JPMJPF2014), JST Moonshot R\&D (Grant
Nos.\ JPMJMS256L and JPMJMS256F), MEXT Q-LEAP (Grant Nos.\ JPMXS0118068682 and JPMXS0120319794), JST
SPRING (Grant No.\ JPMJSP2138), and JST ASPIRE (Grant No.\ JPMJAP2513).

The authors acknowledge the use of GPT-6 Astra (OpenAI) and Claude Fable 5.1 (Anthropic) to assist
in preparing the manuscript, reviewing the literature, and developing experimental scripts. The
authors reviewed and verified the AI-assisted content and take full responsibility for the final
manuscript.
\end{acknowledgments}

\section*{Data availability}
The data and analysis code supporting this work are openly available on
Zenodo~\cite{machino2026data}.

\appendix

\section{Effective interactions}
\label{app:model_details}

\subsection{\xISWAP{}, \xBSWAP{}, and CR}
\label{app:unified_interactions}
\label{app:cr_assignment}
\label{app:two_level_matrix}

The dressed-state matrix $M$ provides a common description of \xISWAP{}, \xBSWAP{}, and CR. We take
the main-drive phase as the reference, making the matrix of $H_{\mathrm{A}}$ real. We can then
choose its dressed eigenstates with real components and obtain a real $M$. Projecting
$V_{\mathrm{A}\mathrm{B}}$ onto the dressed computational states of \xQubitA{} and the two lowest
states of \xQubitB{} gives
\begin{equation}
\begin{aligned}
V_{\mathrm{A}\mathrm{B}}
&=\frac{gM_{01}}2(XX+YY)+\frac{gM_{10}}2(XX-YY)\\
&\quad+\frac{g(M_{00}-M_{11})}2ZX
+\frac{g(M_{00}+M_{11})}2IX.
\end{aligned}
\label{eq:coefficients}
\end{equation}
Here the Pauli operators act in the dressed computational basis. The \xISWAP{} term $XX+YY$ couples
$\lvert \tilde0,1\rangle$ and $\lvert \tilde1,0\rangle$ with amplitude $gM_{01}$ and is resonant
when $\tilde E_1^{\mathrm{A}}-\tilde E_0^{\mathrm{A}}=\delta_{\mathrm{B}}$. The \xBSWAP{} term
$XX-YY$ couples $\lvert \tilde0,0\rangle$ and $\lvert \tilde1,1\rangle$ with amplitude $gM_{10}$ and
is resonant when $\tilde E_0^{\mathrm{A}}-\tilde E_1^{\mathrm{A}}=\delta_{\mathrm{B}}$.

For CR, $\omega_{\mathrm d}=\omega_{\mathrm{B}}$ makes the $ZX$ and $IX$ terms resonant. These terms
describe an effective drive on \xQubitB{} induced through its coupling to the driven \xQubitA{}. The
diagonal elements $M_{00}$ and $M_{11}$ set the effective drive on \xQubitB{} for the two states of
the driven control qubit \xQubitA{}. Their difference gives the conditional coefficient
$J_{ZX}=g(M_{00}-M_{11})$, and their sum gives the unconditional drive
$\Omega_{IX}=g(M_{00}+M_{11})$.

At the \xBSWAP{} resonance, the \xISWAP{} term is detuned by $2\delta_{\mathrm{B}}$ and the $ZX$ and
$IX$ terms by $\delta_{\mathrm{B}}$. They can be eliminated perturbatively when $\lvert
gM_{01}\rvert\ll2\delta_{\mathrm{B}}$ and $\lvert gM_{00}\rvert,\lvert
gM_{11}\rvert\ll\delta_{\mathrm{B}}$. Their virtual effects give the second-order energy shifts in
Appendix~\ref{app:diagonal_shifts}.

If \xQubitA{} is an ideal two-level system, its Hamiltonian reads
\begin{equation}
H_{\mathrm{A}}=
\begin{pmatrix}
0&\frac{\Omega_{\mathrm{A}}}{2}\\[2pt]
\frac{\Omega_{\mathrm{A}}}{2}&\delta_{\mathrm{A}}
\end{pmatrix}.
\label{eq:two_level_hamiltonian}
\end{equation}
We consider $\delta_{\mathrm{A}}=\omega_{\mathrm{A}}-\omega_{\mathrm{d}}<0$ and
$\Omega_{\mathrm{A}}\ge0$, corresponding to a drive above the \xQubitA{} frequency. Defining
$\tan\theta=\Omega_{\mathrm{A}}/\lvert\delta_{\mathrm{A}}\rvert$ with $0\le\theta<\tfrac{\pi}{2}$,
the dressed states continuously connected to $\lvert 0\rangle$ and $\lvert 1\rangle$ are
\begin{equation}
\begin{aligned}
\lvert \tilde0\rangle&=\cos\frac{\theta}{2}\lvert 0\rangle+\sin\frac{\theta}{2}\lvert 1\rangle,\\
\lvert \tilde1\rangle&=-\sin\frac{\theta}{2}\lvert 0\rangle+\cos\frac{\theta}{2}\lvert 1\rangle.
\end{aligned}
\label{eq:two_level_states}
\end{equation}
Their energies are $\tilde
E_0^{\mathrm{A}}=\tfrac{1}{2}(\delta_{\mathrm{A}}+\sqrt{\delta_{\mathrm{A}}^2+\Omega_{\mathrm{A}}^2})$
and $\tilde
E_1^{\mathrm{A}}=\tfrac{1}{2}(\delta_{\mathrm{A}}-\sqrt{\delta_{\mathrm{A}}^2+\Omega_{\mathrm{A}}^2})$.
The state labels are assigned by continuity from the undriven states. They do not indicate
increasing energy in the rotating frame.

Substituting these real eigenstates and $a=\lvert0\rangle\langle1\rvert$ into
Eq.~\eqref{eq:driven_dressing} gives the dressed-state matrix in Eq.~\eqref{eq:two_level_matrix},
including the sign of its reverse element $M_{10}$.
The interaction therefore takes the explicit form
\begin{equation}
\begin{aligned}
V_{\mathrm{A}\mathrm{B}}
&=\frac{g(1+\cos\theta)}4(XX+YY)\\
&\quad-\frac{g(1-\cos\theta)}4(XX-YY)
+\frac{g\sin\theta}2ZX.
\end{aligned}
\label{eq:two_level_coefficients}
\end{equation}
The three terms correspond to \xISWAP{}, \xBSWAP{}, and CR, respectively. The \xBSWAP{} and CR
coefficients are
\begin{equation}
J_{XY}=g(1-\cos\theta),\quad J_{ZX}=g\sin\theta.
\label{eq:two_level_interaction_rates}
\end{equation}
The unconditional drive vanishes because $M_{00}=-M_{11}$. At zero drive, only the $XX+YY$ term
remains. The resonance conditions select which interaction is active.

In the Duffing model at $\omega_{\mathrm d}=\omega_{\mathrm{B}}$, first-order perturbation theory in
$\Omega_{\mathrm{A}}$ recovers the weak-drive result
$J_{ZX}=g\Omega_{\mathrm{A}}\lvert\alpha_{\mathrm{A}}\rvert/[\Delta(\Delta+\lvert\alpha_{\mathrm{A}}\rvert)]$.
This approximation is insufficient at strong drive~\cite{li2022nonperturbative}, where we evaluate
the coefficients directly from the multilevel dressed
eigenstates~\cite{tripathi2019operation,malekakhlagh2020first}.

Driving the higher-frequency qubit can lead to frequency collisions with its own higher-level
transitions~\cite{heya2024floquet,inoue2026systematic}. We therefore consider driving only the
lower-frequency qubit for both \xBSWAP{} and CR.

\subsection{Second-order diagonal shifts}
\label{app:diagonal_shifts}

In the Duffing model, the energy of the $m$th undriven level of \xQubitB{} is given by
$E_m^{\mathrm{B}}=m\delta_{\mathrm{B}}+\alpha_{\mathrm{B}} m(m-1)/2$. At the \xBSWAP{} resonance,
off-resonant couplings identified in Appendix~\ref{app:unified_interactions} and virtual transitions
through higher transmon levels produce second-order energy shifts,
\begin{equation}
E_{km}^{(2)}=\mathop{\sum\nolimits^{\prime}}\limits_{l,n}
\frac{\lvert\langle\tilde l,n\rvert V_{\mathrm{A}\mathrm{B}}\lvert\tilde k,m\rangle\rvert^2}
{\tilde{E}_{k}^{\mathrm{A}}+E_m^{\mathrm{B}}-\tilde{E}_{l}^{\mathrm{A}}-E_n^{\mathrm{B}}}.
\label{eq:second_order_shifts}
\end{equation}
The prime excludes the reference state and its resonant partners. Their resonant coupling is treated
at first order. For \xQubitB{} in $\lvert m\rangle$, the shift in the \xQubitA{} transition energy
is $E_{1m}^{(2)}-E_{0m}^{(2)}$. The difference between these shifts for $m=1$ and $m=0$ gives the
conditional frequency shift $\zeta$:
\begin{equation}
\zeta=E_{11}^{(2)}+E_{00}^{(2)}
-E_{01}^{(2)}-E_{10}^{(2)}.
\label{eq:second_order_zz}
\end{equation}

Thus, the $XX-YY$ and \xZZ{} interactions arise at first and second order in $g$, respectively,
while both depend nonperturbatively on the drive amplitude through the dressed states.

The same corrected energies determine the local detunings:
\begin{equation}
\begin{aligned}
E_{km}&=\tilde{E}_k^{\mathrm{A}}+E_m^{\mathrm{B}}+E_{km}^{(2)},\\
\tilde{\delta}_{\mathrm{A}}&=\frac{E_{10}+E_{11}-E_{00}-E_{01}}{2},\\
\tilde{\delta}_{\mathrm{B}}&=\frac{E_{01}+E_{11}-E_{00}-E_{10}}{2}.
\end{aligned}
\label{eq:local_coefficients}
\end{equation}
Here $\tilde{\delta}_{\mathrm{A}}$ and $\tilde{\delta}_{\mathrm{B}}$ include dressing and
coupling-induced energy shifts and reduce to $\delta_{\mathrm{A}}$ and $\delta_{\mathrm{B}}$ in the
absence of drive and coupling. The \xZZ{} term captures the part that cannot be written as a sum of
single-qubit energies. Within the dressed computational subspace, dropping the common energy offset
gives the effective Hamiltonian
\begin{equation}
\begin{aligned}
H_{\mathrm{eff}}
&=-\frac{J_{XY}}{4}(XX-YY)+\frac{\zeta}{4}\xZZ{}\\
&\quad-\frac{\tilde{\delta}_{\mathrm{A}}}{2}ZI-\frac{\tilde{\delta}_{\mathrm{B}}}{2}IZ.
\end{aligned}
\label{eq:heff_with_shifts}
\end{equation}
After including these energy shifts, resonance occurs when
$\tilde\delta_{\mathrm{A}}+\tilde\delta_{\mathrm{B}}=0$. The $ZI$ and $IZ$ terms produce
single-qubit phase shifts. These shifts are compensated by virtual-$Z$
updates~(Appendix~\ref{app:compiler}).

Without a drive, the \xZZ{} interaction coefficient reduces to
\begin{equation}
\zeta_0=2g^2\left(\frac{1}{\Delta+\lvert\alpha_{\mathrm{A}}\rvert}
-\frac{1}{\Delta-\lvert\alpha_{\mathrm{B}}\rvert}\right).
\label{eq:static_zz_theory}
\end{equation}
Strong dressing can change both the magnitude and sign of the \xZZ{} coefficient. Within the Duffing
model, using the illustrative parameters in Fig.~\ref{fig:resonance_comparison} and
$\xOverTwoPi{\alpha_{\mathrm{B}}}=-250\,\mathrm{MHz}$, the static value is
$\xOverTwoPi{\zeta_0}\simeq-116\,\mathrm{kHz}$. At a drive frequency of $4.6\,\mathrm{GHz}$, the
resonant amplitude $\xOverTwoPi{\Omega_{\mathrm{A}}}\simeq536\,\mathrm{MHz}$ instead gives
$\xOverTwoPi{\zeta}\simeq120\,\mathrm{kHz}$ in the second-order model.

\subsection{siZZle response}
\label{app:sizzle_derivation}

The second tone in Sec.~\ref{sec:sizzle_theory} adds the term
$\frac{\Omega_{\mathrm{B}}}{2}(e^{\mathrm{i}\varphi}b^{\dagger}+e^{-\mathrm{i}\varphi}b)$ to
$H_{\mathrm{B}}$. Keeping the strong-drive dressing of \xQubitA{} in $M^{\mathrm{A}}$, we treat the
weak drive on \xQubitB{} to first order in $\Omega_{\mathrm{B}}$. This drive mixes neighboring
levels of \xQubitB{}, so the expectation values $M^{\mathrm{B}}_{mm}=\langle\tilde m\rvert
b\lvert\tilde m\rangle$ become nonzero. For the dressed states connected to $\lvert 0\rangle$ and
$\lvert 1\rangle$, the Duffing model with $\lambda_2^{\mathrm{B}}=\sqrt2$ gives
\begin{equation}
\begin{aligned}
M^{\mathrm{B}}_{00}&\simeq-\frac{\Omega_{\mathrm{B}} e^{\mathrm{i}\varphi}}{2\delta_{\mathrm{B}}},\\
M^{\mathrm{B}}_{11}&\simeq\frac{\Omega_{\mathrm{B}} e^{\mathrm{i}\varphi}}{2}
\left(\frac{1}{\delta_{\mathrm{B}}}-\frac{2}{\delta_{\mathrm{B}}+\alpha_{\mathrm{B}}}\right).
\end{aligned}
\label{eq:b_dipoles}
\end{equation}
The two terms in $M^{\mathrm{B}}_{11}$ come from the weak drive mixing $\lvert 1\rangle$ with
$\lvert 0\rangle$ and $\lvert 2\rangle$, respectively.

Using the main-drive phase reference, we choose the dressed eigenstates of \xQubitA{} so that
$M^{\mathrm{A}}$ is real. To first order in $g$, the diagonal energy correction is the expectation
value of $V_{\mathrm{A}\mathrm{B}}$ in the product dressed state $\lvert \tilde k,\tilde m\rangle$:
\begin{equation}
\begin{aligned}
E_{km}^{(1)}
&=\langle\tilde k,\tilde m\rvert V_{\mathrm{A}\mathrm{B}}\lvert\tilde k,\tilde m\rangle\\
&=gM^{\mathrm{A}}_{kk}\bigl(M^{\mathrm{B}}_{mm}+M^{\mathrm{B} *}_{mm}\bigr)\\
&=2gM^{\mathrm{A}}_{kk}\operatorname{Re}M^{\mathrm{B}}_{mm}.
\end{aligned}
\label{eq:first_order_diagonal_shifts}
\end{equation}
For \xQubitB{} in $\lvert \tilde m\rangle$, the corresponding shift in the \xQubitA{} transition
energy is $E_{1m}^{(1)}-E_{0m}^{(1)}$. The difference between these shifts for $m=1$ and $m=0$ gives
the additional \xZZ{} contribution:
\begin{equation}
\begin{aligned}
\zeta(\Omega_{\mathrm{B}},\varphi)-\zeta(0)
&\simeq2g(M^{\mathrm{A}}_{11}-M^{\mathrm{A}}_{00})
\operatorname{Re}\bigl(M^{\mathrm{B}}_{11}-M^{\mathrm{B}}_{00}\bigr)\\
&=\frac{2g\Omega_{\mathrm{B}}\alpha_{\mathrm{B}}(M^{\mathrm{A}}_{11}-M^{\mathrm{A}}_{00})}
{\delta_{\mathrm{B}}(\delta_{\mathrm{B}}+\alpha_{\mathrm{B}})}\cos\varphi,
\end{aligned}
\label{eq:sizzle_dipole_response}
\end{equation}
where $\zeta(0)$ is the single-drive baseline and $\zeta(\Omega_{\mathrm{B}},\varphi)$ is the \xZZ{}
interaction coefficient under the two-tone drive. Comparison of
Eqs.~\eqref{eq:sizzle_dipole_response} and~\eqref{eq:sizzle_response} gives
\begin{equation}
\chi_{ZZ}=\frac{2g\alpha_{\mathrm{B}}(M^{\mathrm{A}}_{11}-M^{\mathrm{A}}_{00})}
{\delta_{\mathrm{B}}(\delta_{\mathrm{B}}+\alpha_{\mathrm{B}})}.
\label{eq:sizzle_coefficient}
\end{equation}
The leading-order cancellation condition is
\begin{equation}
\Omega_{\mathrm{B}}\cos\varphi=-\frac{\zeta(0)}{\chi_{ZZ}},
\label{eq:sizzle_null}
\end{equation}
provided that $\chi_{ZZ}\ne0$ and the required amplitude remains in the weak-tone regime. For the
illustrative operating point in Appendix~\ref{app:diagonal_shifts}, $\chi_{ZZ}\simeq0.024$ gives
cancellation at $\varphi=\pi$ and $\xOverTwoPi{\Omega_{\mathrm{B}}}\simeq5\,\mathrm{MHz}$ with the
main drive fixed. Thus a weak additional tone can compensate the \xZZ{} interaction generated by a
much stronger \xBSWAP{} drive.

If the drive on \xQubitA{} is also weak compared with its detunings from the $\lvert
0\rangle\mathord{\leftrightarrow}\lvert 1\rangle$ and $\lvert
1\rangle\mathord{\leftrightarrow}\lvert 2\rangle$ transitions, the first-order result for \xQubitA{}
(with $\varphi=0$) is
$M^{\mathrm{A}}_{11}-M^{\mathrm{A}}_{00}\simeq\Omega_{\mathrm{A}}\alpha_{\mathrm{A}}/[\delta_{\mathrm{A}}(\delta_{\mathrm{A}}+\alpha_{\mathrm{A}})]$.
Substituting this difference into Eq.~\eqref{eq:sizzle_response} and replacing the baseline
$\zeta(0)$ with the static value $\zeta_0$, which neglects drive-dependent corrections of second and
higher order in $g$, gives the usual weak-drive siZZle formula~\cite{wei2022hamiltonian},
\begin{equation}
\zeta\simeq\zeta_0+
\frac{2g\Omega_{\mathrm{A}}\Omega_{\mathrm{B}}\alpha_{\mathrm{A}}\alpha_{\mathrm{B}}\cos\varphi}
{\delta_{\mathrm{A}}\delta_{\mathrm{B}}(\delta_{\mathrm{A}}+\alpha_{\mathrm{A}})(\delta_{\mathrm{B}}+\alpha_{\mathrm{B}})}.
\label{eq:standard_weak_sizzle}
\end{equation}

When the drive on \xQubitB{} is no longer weak, both single-qubit Hamiltonians, including the second
tone, are diagonalized numerically. We evaluate $M^{\mathrm{B}}$ in the dressed states of \xQubitB{}
with the drive phase $\varphi$ included in $H_{\mathrm{B}}$. The diagonal elements of
$V_{\mathrm{A}\mathrm{B}}$ then generate a \xZZ{} contribution at first order in $g$,
\begin{equation}
\zeta^{(1)}=2g\operatorname{Re}\!\left[
(M^{\mathrm{A}}_{11}-M^{\mathrm{A}}_{00})(M^{\mathrm{B}}_{11}-M^{\mathrm{B}}_{00})^*\right],
\label{eq:both_driven_zz_first}
\end{equation}
The second-order shifts of Eq.~\eqref{eq:second_order_shifts}, evaluated with the product dressed
states, are added to this contribution. For $\Omega_{\mathrm{B}}=0$ the first-order term vanishes
and the baseline $\zeta(0)$ is recovered. In the Duffing model, the terms linear in
$\Omega_{\mathrm{B}}$ recover the contribution $\chi_{ZZ}\Omega_{\mathrm{B}}\cos\varphi$ to
$\zeta(\Omega_{\mathrm{B}},\varphi)$ in Eq.~\eqref{eq:sizzle_response}.

The exchange matrix element becomes
\begin{equation}
\langle\tilde0,\tilde0\rvert V_{\mathrm{A}\mathrm{B}}\lvert\tilde1,\tilde1\rangle
=g\bigl[M^{\mathrm{A}}_{10}M^{\mathrm{B}}_{01}
+M^{\mathrm{A}}_{01}M^{\mathrm{B} *}_{10}\bigr].
\label{eq:both_driven_exchange}
\end{equation}
The magnitude and phase of this matrix element determine the exchange strength and axis,
respectively. The resonance, exchange coefficient, and \xZZ{} shift must therefore be evaluated
together when \xQubitB{} is driven strongly.

\subsection{\xBSWAP{} and CR at strong drive}
\label{app:duffing_asymptotics}

In the Duffing model, $E_m^i=m\delta_i+\alpha_i m(m-1)/2$ and $\lambda_m^i=\sqrt m$. The operators
$a$ and $b$ therefore obey the bosonic commutation relations before level truncation. A
displaced-oscillator approximation gives analytical estimates of the \xBSWAP{} and CR coefficients
at strong drive. We compare these estimates with numerical diagonalization of the full
time-independent Duffing Hamiltonian within a finite level truncation. Both treatments use the drive
rotating-wave approximation and keep the interqubit coupling only to first order in $g$.

Writing $a=a'+\beta$ with real displacement $\beta>0$ gives
\begin{equation}
\begin{aligned}
H_{\mathrm{A}}&=\delta_{\mathrm{A}}\,a^{\dagger}a
+\frac{\alpha_{\mathrm{A}}}{2}a^{\dagger2}a^{2}\\
&\quad+\frac{\Omega_{\mathrm{A}}}{2}\left(a+a^{\dagger}\right)\\
&=\mathrm{const.}
+\left(\delta_{\mathrm{A}}\beta+\alpha_{\mathrm{A}}\beta^{3}
+\frac{\Omega_{\mathrm{A}}}{2}\right)(a'+a'^{\dagger})\\
&\quad+\left(\delta_{\mathrm{A}}+2\alpha_{\mathrm{A}}\beta^{2}\right)a'^{\dagger}a'\\
&\quad+\frac{\alpha_{\mathrm{A}}\beta^{2}}{2}
\left(a'^{2}+a'^{\dagger2}\right)+H_3+H_4,
\end{aligned}
\label{eq:duffing_expansion}
\end{equation}
where $H_3$ and $H_4$ collect the cubic and quartic terms. The two quadratic terms define $H_2$. The
displacement is chosen to remove the linear term,
\begin{equation}
\Omega_{\mathrm{A}}=-2\beta\left(\delta_{\mathrm{A}}+\alpha_{\mathrm{A}}\beta^{2}\right).
\label{eq:duffing_displacement}
\end{equation}

To diagonalize $H_2$, we further apply a Bogoliubov transformation to the displaced operator $a'$,
obtaining the annihilation operator $\tilde a$:
\begin{equation}
a'=u\,\tilde a-v\,\tilde a^{\dagger},
\quad u=\cosh\gamma,
\quad v=\sinh\gamma.
\label{eq:duffing_bogoliubov}
\end{equation}
The relation $u^{2}-v^{2}=1$ preserves the bosonic commutator. We choose $\gamma$ to remove the
terms proportional to $\tilde a^2$ and $\tilde a^{\dagger 2}$, giving
\begin{equation}
\tanh(2\gamma)
=\frac{\alpha_{\mathrm{A}}\beta^{2}}
{\delta_{\mathrm{A}}+2\alpha_{\mathrm{A}}\beta^{2}}.
\label{eq:duffing_squeezing_parameter}
\end{equation}
Within this quadratic approximation, the transformed Hamiltonian is
\begin{equation}
H_2\simeq(\widetilde{\omega}_{\mathrm{A}}-\omega_{\mathrm d})\tilde a^{\dagger}\tilde
a+\mathrm{const.},
\label{eq:duffing_quadratic_hamiltonian}
\end{equation}
where $\widetilde{\omega}_{\mathrm{A}}$ is the dressed transition frequency in the lab frame defined
in Eq.~\eqref{eq:dressed_transition_frequency}. For $\delta_{\mathrm{A}}<0$ and
$\alpha_{\mathrm{A}}<0$, the rotating-frame coefficient is negative. Its positive magnitude is
\begin{equation}
\omega_{\mathrm d}-\widetilde{\omega}_{\mathrm{A}}
\simeq\sqrt{
\left(\delta_{\mathrm{A}}+\alpha_{\mathrm{A}}\beta^{2}\right)
\left(\delta_{\mathrm{A}}+3\alpha_{\mathrm{A}}\beta^{2}\right)}.
\label{eq:duffing_fluctuation_gap}
\end{equation}
The cubic and quartic terms $H_3$ and $H_4$ are treated perturbatively below. The numerical
calculation diagonalizes the full Hamiltonian in Eq.~\eqref{eq:duffing_expansion}, including these
terms.

Within the quadratic approximation, the dressed states are displaced squeezed Fock states, $\lvert
\tilde k\rangle\simeq D(\beta)S(\gamma)\lvert k\rangle$, where $D$ and $S$ are the displacement and
squeezing operators. Equation~\eqref{eq:duffing_bogoliubov} gives $M_{kk}\simeq\beta$, $M_{01}\simeq
u$, and $M_{10}\simeq-v$. The common diagonal element gives $\Omega_{IX}\simeq2g\beta$. The
squeezing-induced reverse matrix element gives $J_{XY}\simeq2gv$. At this order
$M_{00}-M_{11}\simeq0$. Thus, the conditional CR coefficient first appears through state-dependent
corrections.

At the endpoint of the \xBSWAP{} resonance between the qubit frequencies, $\delta_{\mathrm{A}}$
approaches zero. First-order perturbation in $H_4$ and second-order perturbation in $H_3$ shift the
positive gap by a total of $-\lvert\alpha_{\mathrm{A}}\rvert/6$. Combining this correction with
Eq.~\eqref{eq:duffing_displacement} gives
\begin{equation}
\begin{aligned}
\beta_{\mathrm{end}}^{2}
&\simeq
\frac{\Delta+\lvert\alpha_{\mathrm{A}}\rvert/6}
{\sqrt{3}\lvert\alpha_{\mathrm{A}}\rvert},\\
\Omega_{\mathrm{A}}^{\mathrm{end}}
&\simeq
\frac{2}{3^{3/4}}
\frac{\left(\Delta+\lvert\alpha_{\mathrm{A}}\rvert/6\right)^{3/2}}
{\lvert\alpha_{\mathrm{A}}\rvert^{1/2}}.
\end{aligned}
\label{eq:duffing_endpoint_amplitude}
\end{equation}
The quadratic squeezing parameter at the endpoint obeys
\begin{equation}
\tanh(2\gamma)=\frac{1}{2},
\quad
\cosh(2\gamma)=\frac{2}{\sqrt{3}},
\quad
v^{2}=\frac{1}{\sqrt{3}}-\frac{1}{2},
\label{eq:duffing_endpoint_squeezing}
\end{equation}
and therefore
\begin{equation}
\frac{J_{XY}^{\mathrm{end}}}{g}
=2v
=\sqrt{\frac{4}{\sqrt{3}}-2}
\simeq0.556.
\label{eq:duffing_endpoint_exchange}
\end{equation}
Perturbing the normalized dressed states through second order gives
\begin{equation}
\begin{aligned}
\epsilon
&=\frac{\lvert\alpha_{\mathrm{A}}\rvert}
{\Delta+\lvert\alpha_{\mathrm{A}}\rvert/6},\\
\frac{J_{XY}^{\mathrm{end}}}{g}
&\simeq0.556238+0.129670\epsilon+0.033816\epsilon^{2}.
\end{aligned}
\label{eq:duffing_endpoint_correction}
\end{equation}
For the parameters of Fig.~\ref{fig:resonance_comparison}, these perturbative expressions give
$\Omega_{\mathrm{A}}^{\mathrm{end}}/\Delta=1.6481$ and $J_{XY}^{\mathrm{end}}/g=0.6006$. Direct
numerical diagonalization of the full Duffing Hamiltonian gives $1.6489$ and $0.6011$, respectively.

For CR, the drive frequency is fixed at $\omega_{\mathrm d}=\omega_{\mathrm{B}}$. Hence,
$\delta_{\mathrm{A}}=-\Delta$. The Hellmann--Feynman theorem states that the derivative of a dressed
energy with respect to the drive amplitude equals the expectation value of the corresponding drive
term. The difference between the two dressed states therefore gives
\begin{equation}
\frac{J_{ZX}}{g}
=\frac{\partial(\tilde E_0^{\mathrm{A}}-\tilde E_1^{\mathrm{A}})}
{\partial\Omega_{\mathrm{A}}}
=-\frac{\partial\widetilde{\omega}_{\mathrm{A}}}
{\partial\Omega_{\mathrm{A}}}.
\label{eq:cr_hellmann_feynman}
\end{equation}
Substituting the approximate gap of Eq.~\eqref{eq:duffing_fluctuation_gap} into this derivative
relation and defining $z=\lvert\alpha_{\mathrm{A}}\rvert\beta^{2}/\Delta$, we obtain
\begin{equation}
\frac{J_{ZX}}{g}\simeq
\left(\frac{\lvert\alpha_{\mathrm{A}}\rvert}{\Delta}\right)^{1/2}
\frac{z^{1/2}(2+3z)}
{(1+3z)^{3/2}(1+z)^{1/2}}.
\label{eq:cr_asymptotic_rate}
\end{equation}

Maximizing this expression requires $9z^3+12z^2+3z-2=0$, and its positive root $z=0.282$ yields
\begin{equation}
\begin{aligned}
J_{ZX}^{\mathrm{peak}}
&\simeq0.532\,g
\left(\frac{\lvert\alpha_{\mathrm{A}}\rvert}{\Delta}\right)^{1/2},\\
\Omega_{\mathrm{A}}^{\mathrm{peak}}
&\simeq1.36\,
\frac{\Delta^{3/2}}{\lvert\alpha_{\mathrm{A}}\rvert^{1/2}}.
\end{aligned}
\label{eq:cr_asymptotic_maximum}
\end{equation}
For the same illustrative parameters, including the first perturbative correction to the gap gives
$\Omega_{\mathrm{A}}^{\mathrm{peak}}/\Delta=2.7346$ and $J_{ZX}^{\mathrm{peak}}/g=0.3002$. Direct
numerical diagonalization of the Duffing Hamiltonian gives $2.8213$ and $0.2983$, respectively.

\subsection{Full-cosine transmon model}
\label{app:cosine_model}

At zero offset charge, the isolated transmon Hamiltonian is written as $4E_{\mathrm C}\hat
n^2-E_{\mathrm J}\cos\hat\varphi$, where $\hat n$ counts Cooper pairs and $\hat\varphi$ is the
superconducting phase~\cite{koch2007charge,blais2021circuit}. Its eigenenergies $\mathcal E_m$ and
eigenstates $\lvert m\rangle$ define the undriven basis. We choose $E_{\mathrm J}$ and $E_{\mathrm
C}$ to reproduce the specified transition frequency and anharmonicity.

For each transmon, the energies and transition coefficients used in Sec.~\ref{sec:multilevel_model}
are
\begin{equation}
\begin{aligned}
E_m&=\mathcal E_m-\mathcal E_0-m\omega_{\mathrm d},\\
\lambda_{m+1}&=\frac{\langle m\rvert\hat n\lvert m+1\rangle}
{\langle0\rvert\hat n\lvert1\rangle},\quad m\ge0.
\end{aligned}
\end{equation}
Eigenstate phases are chosen to make $\lambda_m$ real and positive, with $\lambda_1=1$. These
coefficients define $a$ and $b$ in Eq.~\eqref{eq:transition_operators}, without assuming bosonic
commutation relations. Retaining adjacent-level transitions in the rotating-wave approximation gives
Eq.~\eqref{eq:multilevel_model}. Thus, $g$ is the exchange matrix element between the undriven
states $\lvert 10\rangle$ and $\lvert 01\rangle$, and $\Omega_{\mathrm{A}}$ retains its resonant
two-level Rabi-frequency normalization.

\section{Device characterization}
\label{app:device_characterization}

\subsection{Drive spectroscopy}
\label{app:wide_spectrum}

\begin{figure}[tbp]
\includegraphics[width=\columnwidth]{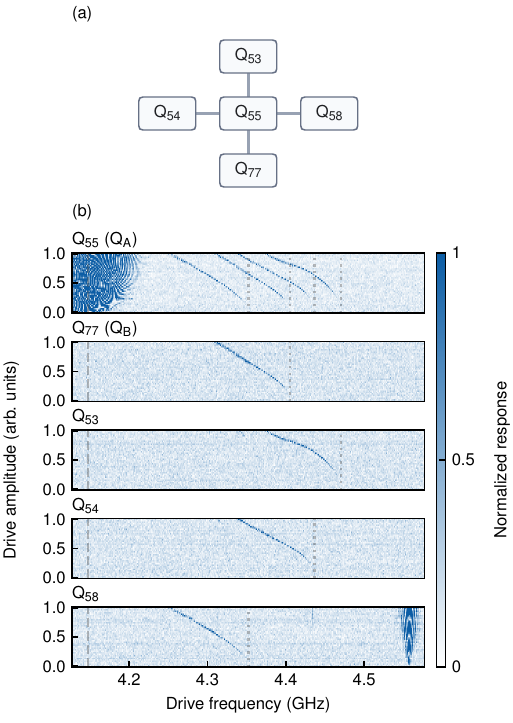}
\caption{Wide-band response to a drive on \xQubitA{}. (a)~Qubit connectivity. (b)~Normalized readout
responses. Dotted lines mark the undriven two-photon midpoints: all four in the \xQubitA{} panel and
the corresponding one in each neighbor panel. Dashed lines mark the \xQubitA{} transition.}
\label{fig:wide_spectrum}
\end{figure}

Figure~\ref{fig:wide_spectrum} extends the frequency--amplitude scan in Fig.~\ref{fig:regime} to the
surrounding qubits. A flat-top probe pulse is applied to \xQubitA{} for a total of
$500\,\mathrm{ns}$, including two $30$\nobreakdash-ns ramps. The color scale shows the magnitude of
the background-subtracted mean IQ response, normalized at each drive amplitude, rather than
population.

In Fig.~\ref{fig:wide_spectrum}(b), the resonance line observed in both the \xQubitA{} and
\xQubitB{} responses corresponds to \xBSWAP{} exchange between $\lvert 00\rangle$ and $\lvert
11\rangle$. The fringes near $\xOverTwoPi{\omega_{\mathrm{A}}}\simeq4.15\,\mathrm{GHz}$ are the
direct one-photon response of \xQubitA{}. A CR response is also visible in $\mathrm{Q}_{58}$ near
its transition frequency of $4.56\,\mathrm{GHz}$.

\begin{figure*}[tbp]
\begin{minipage}[t]{\columnwidth}
\includegraphics[width=\columnwidth]{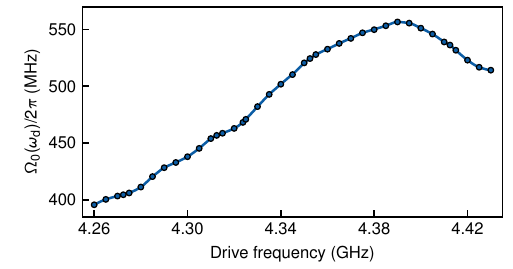}
\caption{Frequency-dependent drive-amplitude scale. Markers show the midpoints of
$\xOverTwoPi{\Omega_0(\omega_{\mathrm{d}})}$ inferred from Stark shifts at amplitude settings
$s=0.15$ and $0.30$. The solid curve is the shape-preserving piecewise cubic interpolation.}
\label{fig:frequency_gain}
\end{minipage}\hfill
\begin{minipage}[t]{\columnwidth}
\includegraphics[width=\columnwidth]{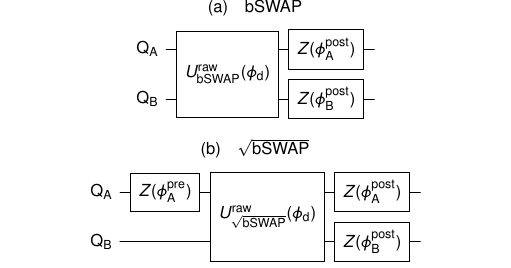}
\caption{Local-phase corrections for the pulse implementations of (a)~\xBSWAP{} and
(b)~$\xSQRTBSWAP$. The superscript `raw' denotes the pulse before local-phase correction.}
\label{fig:phase_compilation}
\end{minipage}
\end{figure*}

\subsection{Drive-amplitude calibration}
\label{app:amplitude_calibration}

To calibrate the amplitude setting of the control electronics, we apply the off-resonant drive
during the Ramsey waiting interval. Comparing the Ramsey oscillation frequencies with and without
the drive gives the Stark shift, from which we infer the physical drive amplitude
$\Omega_{\mathrm{A}}$ using the model in Appendix~\ref{app:cosine_model}.

Writing the normalized amplitude setting as $s\in[0,1]$, we define the frequency-dependent
drive-amplitude scale $\Omega_0(\omega_{\mathrm{d}})$ by
$\Omega_{\mathrm{A}}=s\,\Omega_0(\omega_{\mathrm{d}})$. Stark-Ramsey measurements from $4.260$ to
$4.430\,\mathrm{GHz}$ give the calibration shown in Fig.~\ref{fig:frequency_gain}. The two
weak-drive settings give scales that differ by at most $0.77\%$. At the tested higher drive
amplitudes, the inferred scales are less than $1\%$ below this calibration. The interpolated
calibration is used for the theoretical resonance curves in Fig.~\ref{fig:regime}.

Independent Stark-shift measurements at the \xBSWAP{} and CR drive frequencies give
$\xOverTwoPi{\Omega_{\mathrm{A}}}=418.71$ and $419.08\,\mathrm{MHz}$, respectively, for the
amplitude matching used in Table~\ref{tab:compare}.

\section{Pulse implementation}
\label{app:ramp}
\label{app:compiler}

\subsection{Quasi-adiabatic ramps}

Rapid ramps to the \xBSWAP{} resonance can induce nonadiabatic transitions, whereas slow passage
through intervening resonances can populate unwanted states. For rapid, near-adiabatic transfer
between undriven and target dressed states, we use a pulse that combines a smoothed quasi-adiabatic
ramp with a counter-diabatic
quadrature~\cite{martinis2014fast,berry2009transitionless,gueryodelin2019shortcuts}.

For the two-level reference $H_0(t)=\tfrac12[-\delta_{\mathrm{A}} Z+\Omega_x(t)X]$, define
$\theta(t)=\arctan[\Omega_x(t)/|\delta_{\mathrm{A}}|]$, where
$\delta_{\mathrm{A}}=\omega_{\mathrm{A}}-\omega_{\mathrm d}$. We design the rising ramp, $0\le
t\le\tau$, by prescribing a smooth adiabaticity profile,
\begin{equation}
\epsilon(t)=\frac{|\dot\theta(t)|}{\sqrt{\delta_{\mathrm{A}}^2+\Omega_x(t)^2}}
=\epsilon_{\max}\sin^2\!\left(\frac{\pi t}{\tau}\right).
\label{eq:squad_adiabaticity}
\end{equation}
This profile vanishes smoothly at both ramp boundaries. Integrating with $\Omega_x(0)=0$ and
$\Omega_x(\tau)=\bar\Omega_x$~gives
\begin{equation}
\begin{aligned}
\Omega_x(t)&=|\delta_{\mathrm{A}}|\tan\theta(t),\\
\sin\theta(t)&=\sin\bar\theta
\left[u-\frac{\sin(2\pi u)}{2\pi}\right],\\
\bar\theta&=\arctan\frac{\bar\Omega_x}{|\delta_{\mathrm{A}}|},\quad u=\frac{t}{\tau}.
\end{aligned}
\label{eq:squad}
\end{equation}
The boundary condition gives $\epsilon_{\max}=2\sin\bar\theta/(|\delta_{\mathrm{A}}|\tau)$. The
falling ramp is the time reverse.

The counter-diabatic quadrature contributes $\Omega_y(t)Y/2$ to $H_0(t)$, with
\begin{equation}
\Omega_y(t)=-\frac{\delta_{\mathrm{A}}\dot{\Omega}_x(t)}{\delta_{\mathrm{A}}^2+\Omega_x(t)^2}.
\label{eq:squad_cd}
\end{equation}
The counter-diabatic quadrature has opposite signs on the rising and falling ramps and vanishes on
the plateau.

\subsection{Gate calibration and phase tracking}
\label{sec:calibration_procedure}
\label{app:local_phases}
\label{app:phase_tracking}

At a fixed drive amplitude, frequency--duration chevron patterns provide the initial resonance
frequency and pulse duration. In repeated-pulse scans, we apply identical exchange pulses in
succession and sweep the duration of each pulse while keeping the drive frequency and amplitude
fixed. A cosine fit to the resulting population oscillations refines the pulse duration.

To calibrate the phases, we apply each gate to $\lvert 0+\rangle$, $\lvert 1+\rangle$, $\lvert
+0\rangle$, and $\lvert +1\rangle$, where $\lvert +\rangle=(\lvert 0\rangle+\lvert
1\rangle)/\sqrt2$, and reconstruct the output states by two-qubit tomography. With the exchange
angle fixed, a joint fit to the off-diagonal density-matrix elements determines the local phases and
residual \xZZ{} phase.

The independent residual-phase checks in Sec.~\ref{sec:zz} average the estimates from coherence on
the two qubits. Their $95\%$ statistical intervals, obtained from multinomial counting statistics,
are $[2.2,6.4]\times10^{-3}\,\mathrm{rad}$ for \xBSWAP{} and $[-0.3,4.9]\times10^{-3}\,\mathrm{rad}$
for $\xSQRTBSWAP$.

Figure~\ref{fig:phase_compilation} uses the local-phase decomposition described in
Ref.~\citenum{wei2024native}: two post-rotations for \xBSWAP{} and an additional pre-rotation for
$\xSQRTBSWAP$. The pre-rotation in Fig.~\ref{fig:phase_compilation}(b) is implemented through the
\xBSWAP{} drive phase and virtual post-rotations, without physical $Z$ pulses. The required
drive-phase adjustment is half the pre-$Z$ rotation angle because of the two-photon transition
process.

\section{Interaction anisotropy}
\label{app:anisotropy}
\label{app:zz}

The exchange and \xZZ{} coefficients directly determine the interaction anisotropy. An $X$ basis
change on \xQubitA{} puts the interaction in the \xXXZ{} form,
\begin{equation}
\begin{aligned}
H'_{\mathrm{int}}&=-\frac14\left[J_{XY}(XX+YY)+\zeta\xZZ{}\right],\\
\Delta_{XXZ}&=\frac{\zeta}{J_{XY}}.
\end{aligned}
\label{eq:xxz_anisotropy}
\end{equation}
Here, $\Delta_{XXZ}$ is the ratio of the coefficient of \xZZ{} to that of $XX+YY$. In this
transformed basis, $\zeta=0$ gives an $XX+YY$ interaction, while $\zeta=J_{XY}$ gives the isotropic
Heisenberg interaction.

For a finite resonant pulse with local phases removed, we compare the integrated coefficients,
\begin{equation}
\Delta_{\mathrm{eff}}=\frac{\int_0^T\zeta(t)\,\mathrm{d}t}{\int_0^T J_{XY}(t)\,\mathrm{d}t}.
\label{eq:integrated_anisotropy}
\end{equation}
For the calibrated \xBSWAP{} and $\xSQRTBSWAP$ pulses, this gives
$\Delta_{\mathrm{eff}}=4\Phi_{ZZ}/\pi$ and $8\Phi_{ZZ}/\pi$, respectively. The measured phases in
Fig.~\ref{fig:q55_zz_controls} therefore correspond to sampled $\Delta_{\mathrm{eff}}$ ranges of
approximately $-0.09$ to $0.21$ for \xBSWAP{} and $-0.12$ to $0.18$ for $\xSQRTBSWAP$. Both include
positive and negative anisotropy but remain below the isotropic point, $\Delta_{\mathrm{eff}}=1$.
Extending this range toward the isotropic point would require stronger siZZle tones and joint
calibration of the siZZle and \xBSWAP{} drives.

\section{Benchmarking protocols}
\label{app:benchmarks}

\subsection{Interleaved randomized benchmarking}
\label{app:irb_statistics}

We benchmark the native \xBSWAP{} using standard
IRB~\cite{magesan2012efficient,magesan2012characterizing}. The reference and interleaved decay
parameters, $p_{\mathrm{ref}}$ and $p_{\mathrm{int}}$, give the gate error under the depolarizing
model,
\begin{equation}
r_{\mathrm{gate}}=\frac34\left(1-\frac{p_{\mathrm{int}}}{p_{\mathrm{ref}}}\right).
\label{eq:depolarizing_gate_error}
\end{equation}
The data in Fig.~\ref{fig:gate_benchmarks}(a) are mean populations $P(00)$ over $32$ random
sequences at each depth, with one-standard-error bars. The reference and interleaved means are
fitted separately to exponentials with free offsets.

The reference Cliffords use \xBSWAP{} and single-qubit gates, with one \xBSWAP{} as the interleaved
operation.

\label{app:sqrt_pair_irb}

\begin{figure*}[t]
\begin{minipage}[t]{\columnwidth}
\includegraphics[width=\columnwidth]{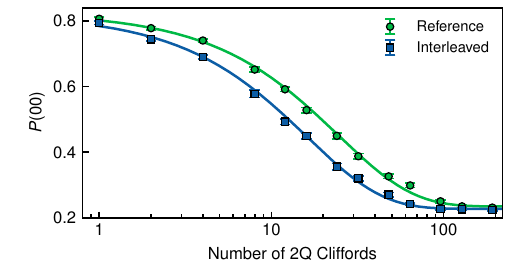}
\caption{Interleaved randomized benchmarking of \xBSWAP{} composed of two $\xSQRTBSWAP$ pulses.
Markers show mean populations with one-standard-error bars. Lines are separate fits of the reference
and interleaved means to an exponential decay with an offset.}
\label{fig:sqrt_pair_irb}
\end{minipage}\hfill
\begin{minipage}[t]{\columnwidth}
\includegraphics[width=\columnwidth]{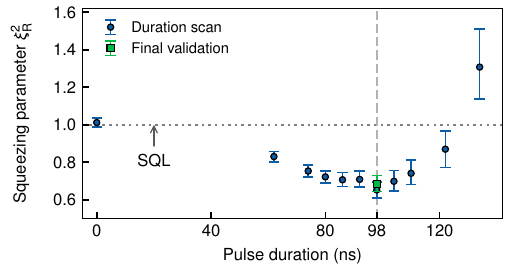}
\caption{Preparation-time dependence of two-qubit spin squeezing. Blue circles show
\xSqueezingParameter{} along the squeezed axis determined separately at each pulse duration. The
green square shows the independent validation reported in Sec.~\ref{sec:squeezing}. Error bars show
$95\%$ statistical intervals. The horizontal dotted line marks the SQL, and the vertical dashed line
marks $98\,\mathrm{ns}$.}
\label{fig:squeezing_duration}
\end{minipage}
\end{figure*}

For $\xSQRTBSWAP$, we instead benchmark a pair of pulses. Two consecutive $150$-ns pulses with the
same calibrated parameters form a $300$-ns \xBSWAP{} block. This block is used in both the reference
Clifford compilation and the interleaved operation~(Fig.~\ref{fig:sqrt_pair_irb}).

The block error is $r_2=1.54\times10^{-2}$. Assuming small additive errors for the two
pulses~\cite{wei2024native}, the per-pulse estimate is $r_1^{\mathrm{est}}=r_2/2=7.70\times10^{-3}$.
This gives an estimated fidelity of $99.23\%$, with a $95\%$ statistical interval of
$[99.08,99.40]\%$. This is a block-derived estimate rather than a direct single-pulse IRB
measurement. The XEB and IRB-derived estimates are consistent within their reported statistical
uncertainties, with overlapping $95\%$ intervals.

\subsection{Interleaved XEB with single-qubit references}

Interleaved XEB compares reference circuits containing random single-qubit
layers~\cite{boixo2018characterizing,arute2019quantum} with circuits that insert a $150$-ns
$\xSQRTBSWAP$ between the layers~\cite{zhang2024tunable}. The reference omits two-qubit pulses
without replacing them by idle periods. We evaluate both circuit sets using regression XEB
scores~\cite{arute2019quantum}. The data in Fig.~\ref{fig:gate_benchmarks}(b) use $64$ random
circuits at each depth, with one-standard-error bars estimated by circuit bootstrap.

The reference XEB score combines the decay of each qubit with their product and therefore is not, in
general, a single exponential~\cite{simakov2026decay}. We fit each qubit's exponential decay
separately to obtain $\lambda_{\mathrm{A}}$ and $\lambda_{\mathrm{B}}$. Their equivalent two-qubit
depolarizing parameter is
\begin{equation}
p_{\mathrm{ref}}=\frac{\lambda_{\mathrm{A}}+\lambda_{\mathrm{B}}+3\lambda_{\mathrm{A}}\lambda_{\mathrm{B}}}{5}.
\label{eq:xeb_reference_parameter}
\end{equation}
The interleaved score gives $p_{\mathrm{int}}$ from an exponential fit with a free offset. If the
reference and interleaved parameters admit the same two-qubit depolarizing description,
Eq.~\eqref{eq:depolarizing_gate_error} gives the gate error~\cite{zhang2024tunable}. The
\xZZ{}-compensated data in Fig.~\ref{fig:gate_benchmarks}(b) give
$\mathcal{F}=1-r_{\mathrm{gate}}=99.34\%$, with a $95\%$ interval $[99.27,99.42]\%$.

\section{Spin-squeezing measurements}
\label{app:squeezing_preparation}

Starting from $\lvert 00\rangle$, we prepare the state in Fig.~\ref{fig:squeezing} using a $98$-ns
pulse at $4.312\,\mathrm{GHz}$, including two $24$-ns ramps. Population oscillations indicate a
rotation angle of approximately $\pi/4$ in the even-parity subspace.

We quantify squeezing using the Wineland parameter~\cite{wineland1994squeezed}
\begin{equation}
\xSqueezingParameter{}(\phi)=\frac{2\operatorname{Var}(S_\phi)}{\langle S_z\rangle^2},
\label{eq:spin_squeezing}
\end{equation}
where $S_\phi=S_x\cos\phi+S_y\sin\phi$ is the transverse spin component at angle $\phi$. Values
below the coherent-state SQL of $\xSqueezingParameter{}=1$ indicate spin squeezing. We measure the
transverse spin fluctuations by applying collective rotations before population
readout~\cite{meyer2001experimental}.

We scan the preparation-pulse duration at fixed drive frequency, amplitude, and ramp duration. At
each duration, a sinusoidal fit to measurements at $16$ analysis angles identifies the squeezed
axis. Independent measurements of the transverse variance along this axis and the mean longitudinal
spin then determine \xSqueezingParameter{}. The $95\%$ statistical intervals are obtained by
multinomial bootstrap of the measured outcome counts.

Figure~\ref{fig:squeezing_duration} shows the duration scan and the independent validation at
$98\,\mathrm{ns}$, near the observed minimum. Both datasets use initial-state heralding without
readout-error mitigation. The zero-duration point was measured without the preparation pulse. The
increase at longer durations can arise from residual transverse variance due to preparation and
readout errors, amplified by the decreasing $\langle S_z\rangle^2$ in the denominator.

\section{Gate synthesis and comparison with CR}
\label{app:cr_comparison}

\subsection{Gate constructions and ideal times}
\label{app:cnot_cost}
\label{app:composition}

\begin{figure*}[tbp]
\centering
\includegraphics[width=\textwidth]{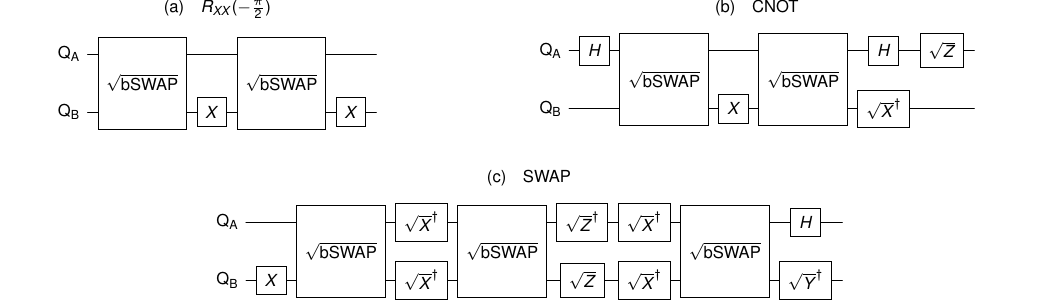}
\caption{Constructions from $\xSQRTBSWAP$ pulses. (a) Echoed $XX$ rotation. (b) \xCNOT{} with
\xQubitA{} as control. (c) \xSWAP{}. Square-root Pauli labels denote positive $\pi/2$ rotations, and
daggers denote inverses.}
\label{fig:gate_applications}
\label{fig:swap_phases}
\end{figure*}

The implementation time depends on both the interaction coefficients and the rotations needed for
the target operation. The coefficients $J_{XY}$ and $J_{ZX}$ in Eqs.~\eqref{eq:multilevel_exchange}
and~\eqref{eq:coefficients} give the interactions
\begin{equation}
\begin{aligned}
H_{\xBSWAP{}}&=-\frac{J_{XY}}4(XX-YY),\\
H_{\mathrm{CR}}&=\frac{J_{ZX}}2ZX.
\end{aligned}
\label{eq:cnot_generators}
\end{equation}
With this normalization, $\xSQRTBSWAP$ and $R_{ZX}(\frac{\pi}{2})$ each generate a maximally
entangled state from a suitable product state in time $\frac{\pi}{2J}$ when $J_{XY}=J_{ZX}=J$. The
coefficients therefore measure entanglement-generation rates on the same scale.

The Cartan decomposition describes a two-qubit gate by rotations along $XX$, $YY$, and
$ZZ$~\cite{zhang2003geometric}. \xCNOT{} and \xCZ{} belong to the same local-equivalence class and
require one axis, whereas \xSWAP{} requires all three.

An $X$ on \xQubitB{} reverses the $YY$ component of $U_{\xSQRTBSWAP}$ while preserving $XX$. Two
native pulses with these echoes therefore give $R_{XX}(-\frac{\pi}{2})$
[Fig.~\ref{fig:gate_applications}(a)]. The echo also reverses \xZZ{}, canceling an identical
residual conditional phase from the two pulses.

An $H$ on \xQubitA{} maps $XX$ to $ZX$. The local rotations in panel (b) then give
\xCNOT{}~\cite{zhang2003geometric}. CR realizes the same local-equivalence class with one
$R_{ZX}(\frac{\pi}{2})$ and local rotations.

The \xSWAP{} construction uses both axes of each native pulse. Local basis changes transform $XX-YY$
into $XX+YY$, $XX+\xZZ{}$, and $YY+\xZZ{}$, giving the three factors in
Fig.~\ref{fig:swap_phases}(c). Since these Pauli products commute,
\begin{equation}
\begin{aligned}
&e^{\mathrm{i}\pi(YY+\xZZ{})/8}
e^{\mathrm{i}\pi(XX+\xZZ{})/8}
e^{\mathrm{i}\pi(XX+YY)/8}\\
&\qquad=e^{\mathrm{i}\pi(XX+YY+\xZZ{})/4}
=e^{\mathrm{i}\pi/4}U_{\xSWAP{}}.
\end{aligned}
\label{eq:swap_synthesis}
\end{equation}
The CR route uses three \xCNOT{} operations, alternating their control directions. These
constructions attain the minimum interaction times with instantaneous local
operations~\cite{vidal2002interaction}. Table~\ref{tab:ideal_times} gives the times for constant
interaction coefficients.

\begin{table}[tbp]
\caption{Ideal interaction times.}
\label{tab:ideal_times}
\begin{ruledtabular}
\renewcommand{\arraystretch}{1.65}
\begin{tabular}{lcc}
Target & CR-based & $\xSQRTBSWAP$-based \\
\colrule
\noalign{\vskip 3pt}
Bell state & $\dfrac{\pi}{2J_{ZX}}$ & $\dfrac{\pi}{2J_{XY}}$ \\[10pt]
\xCNOT{}, \xCZ{} & $\dfrac{\pi}{2J_{ZX}}$ & $\dfrac{\pi}{J_{XY}}$ \\[10pt]
\xSWAP{} & $\dfrac{3\pi}{2J_{ZX}}$ & $\dfrac{3\pi}{2J_{XY}}$ \\
\noalign{\vskip 3pt}
\end{tabular}
\end{ruledtabular}
\end{table}

\subsection{Gate speeds under a common drive limit}
\label{app:speed_bound}

\begin{figure*}[tbp]
\includegraphics[width=\textwidth]{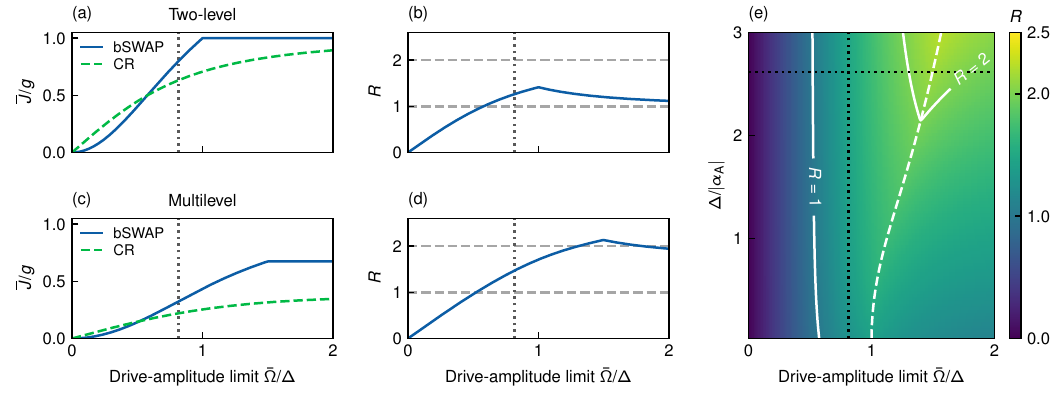}
\caption{Calculated interaction rates under a common drive-amplitude limit. (a) and (b) Two-level
model. (c) and (d) Multilevel model using the experimental device parameters. Panels (a) and (c)
show the maximum interaction coefficients $\bar{J}$ for \xBSWAP{} and CR, normalized by $g$. Panels
(b) and (d) show their ratio $R$. Gray dashed horizontal lines mark the thresholds for Bell-state
preparation and \xSWAP{} ($R=1$), and for \xCNOT{} and \xCZ{} ($R=2$). (e) Multilevel $R$ versus
drive-amplitude limit and detuning $\Delta$ with qubit A fixed, with solid contours at $R=1$ and
$R=2$. The dashed curve in (e) marks the maximum drive amplitude supporting \xBSWAP{} resonance
between the qubit frequencies. The black dotted horizontal line in (e) marks the detuning used in
(c) and (d). Dotted vertical lines mark the common experimental drive amplitude,
$\bar{\Omega}/\Delta\simeq0.81$.}
\label{fig:speed_bound}
\label{fig:speed_domain}
\end{figure*}

\begin{figure}[tbp]
\includegraphics[width=\columnwidth]{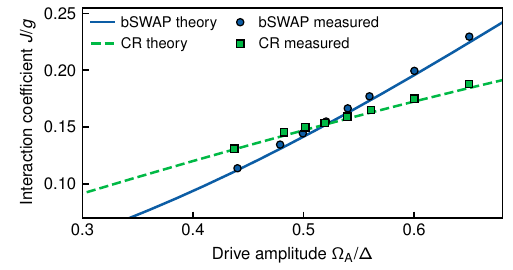}
\caption{Interaction coefficients versus drive amplitude. Circles and squares show measured
\xBSWAP{} and CR coefficients, respectively. Solid blue and dashed green lines show the
corresponding cosine-model predictions.}
\label{fig:measured_interactions}
\end{figure}

The $\xSQRTBSWAP$-based construction is faster than the CR-based construction for Bell-state
preparation and \xSWAP{} when $J_{XY}>J_{ZX}$, whereas \xCNOT{} and \xCZ{} require $J_{XY}>2J_{ZX}$.
To compare the available rates, let $\bar{\Omega}>0$ be the common hardware limit on drive
amplitude. Each protocol maximizes its coefficient over $0\le\Omega_{\mathrm{A}}\le\bar{\Omega}$,
giving $\overline{J_{XY}}$ and $\overline{J_{ZX}}$. For \xBSWAP{}, the drive parameters lie on the
resonance curve between the qubit frequencies, whereas CR uses $\omega_{\mathrm
d}=\omega_{\mathrm{B}}$. Define
\begin{equation}
R(\bar{\Omega})=\frac{\overline{J_{XY}}}{\overline{J_{ZX}}}.
\label{eq:common_drive_frontier}
\end{equation}

Evaluating Eq.~\eqref{eq:two_level_interaction_rates} at the \xBSWAP{} resonance given by
Eq.~\eqref{eq:two_level_resonance} and at $\omega_{\mathrm d}=\omega_{\mathrm{B}}$ for CR gives
\begin{equation}
J_{XY}=\frac{2g\Omega_{\mathrm{A}}^{2}}{\Delta^2+\Omega_{\mathrm{A}}^{2}},\quad
J_{ZX}=\frac{g\Omega_{\mathrm{A}}}{\sqrt{\Delta^2+\Omega_{\mathrm{A}}^{2}}}.
\label{eq:two_level_common_bound}
\end{equation}
The maximum drive amplitude supporting \xBSWAP{} resonance between the qubit frequencies is
$\Omega_{\mathrm{A}}=\Delta$. The largest coefficients therefore use $\min(\bar{\Omega},\Delta)$ for
\xBSWAP{} and $\bar{\Omega}$ for CR. Hence,
\begin{equation}
R(\bar{\Omega})=
\begin{cases}
\dfrac{2\bar{\Omega}}{\sqrt{\Delta^2+\bar{\Omega}^2}},&\bar{\Omega}\le\Delta,\\
\noalign{\vskip 10pt}
\dfrac{\sqrt{\Delta^2+\bar{\Omega}^2}}{\bar{\Omega}},&\bar{\Omega}>\Delta.
\end{cases}
\label{eq:two_level_rate_ratio}
\end{equation}
The maximum value, $R=\sqrt2$, occurs at $\bar{\Omega}=\Delta$. Thus, the $\xSQRTBSWAP$-based
construction can shorten Bell-state preparation and the \xSWAP{} operation for
$\bar{\Omega}>\Delta/\sqrt3$, but not \xCNOT{} or \xCZ{} in this model.

For Fig.~\ref{fig:speed_bound}, we evaluate the same dressed-state matrix $M$ using the model
defined in Appendix~\ref{app:cosine_model}. The calculations use 20 transmon levels. Convergence
checks with 24 levels and a larger charge-basis cutoff at representative parameter points are
included in the accompanying data~\cite{machino2026data}. The parameters $E_{\mathrm J}$ and
$E_{\mathrm C}$ are fixed by the measured transition frequency and anharmonicity in
Table~\ref{tab:device}. Here $R=1.71$ at $\bar{\Omega}=\Delta$ and reaches two at $1.30\Delta$. The
$\xSQRTBSWAP$-based construction can therefore also shorten the ideal \xCNOT{} and \xCZ{}
interaction times at sufficiently strong drive. The maximum drive amplitude supporting the \xBSWAP{}
resonance is $1.50\Delta$, where the resonance frequency approaches $\omega_{\mathrm{A}}$ and
$R\simeq2.13$. Beyond the \xBSWAP{} endpoint, $\overline{J_{XY}}$ remains constant, whereas
$\overline{J_{ZX}}$ continues to increase within the range shown in Fig.~\ref{fig:speed_bound}(c),
reducing $R$.

Figure~\ref{fig:speed_domain}(e) shows the detuning dependence with $\omega_{\mathrm{A}}$ fixed and
$\omega_{\mathrm{B}}$ varied. The $R=1$ boundary lies near $\bar{\Omega}/\Delta=0.52$--$0.57$. The
$R>2$ region requires $\Delta/|\alpha_{\mathrm{A}}|$ above about $2.2$ and drive amplitudes near the
maximum supporting \xBSWAP{} resonance between the qubit frequencies. The \xCNOT{} speed advantage
thus requires both large detuning and high available drive strength.

Figure~\ref{fig:measured_interactions} compares measured interaction coefficients with the
cosine-model predictions shown in Fig.~\ref{fig:speed_domain}(c). We extract $J_{XY}$ from
frequency-resolved even-parity population oscillations and $\sqrt{J_{ZX}^{2}+J_{ZY}^{2}}$ from
Hamiltonian tomography. Drive amplitudes are inferred from Stark shifts as in
Appendix~\ref{app:device_characterization}. The \xBSWAP{} coefficient rises more rapidly with drive
amplitude and overtakes the CR coefficient. Without fitting a scale or offset, the measured
coefficients agree with the model predictions, with root-mean-square deviations in $J/g$ of $0.0036$
for \xBSWAP{} and $0.0023$ for CR.

\subsection{\xCNOT{} and \xSWAP{} implementations}
\label{sec:compare}
\label{app:cr}

The operations in Table~\ref{tab:compare} use drive amplitudes matched at
$\xOverTwoPi{\Omega_{\mathrm{A}}}\simeq420\,\mathrm{MHz}$ by the Stark-shift-based calibration
(Appendix~\ref{app:device_characterization}). Their durations include ramps and physical local
rotations.

The $\xSQRTBSWAP$-based \xCNOT{} and \xSWAP{} use two and three $150$-ns pulses, plus $96$ and
$120\,\mathrm{ns}$ of local operations, respectively. The CR-based implementations use one and three
$268$-ns echoed $R_{ZX}(\frac{\pi}{2})$ pulses, plus $24$ and $48\,\mathrm{ns}$ of local operations.
Figure~\ref{fig:operation_irb_comparison} shows the IRB data underlying the fidelities in
Table~\ref{tab:compare}.

Population oscillations give $\xOverTwoPi{J_{XY}}\simeq1.94\,\mathrm{MHz}$, corresponding to an
ideal $\xSQRTBSWAP$ interaction time of $129\,\mathrm{ns}$ excluding ramp overhead. For the
corresponding device parameters and Stark shift, the model gives
$\xOverTwoPi{\Omega_{\mathrm{A}}}\simeq420\,\mathrm{MHz}$ and
$\xOverTwoPi{J_{XY}}\simeq1.91\,\mathrm{MHz}$.

\raggedbottom
\onecolumngrid
\begin{center}
\includegraphics[width=\textwidth]{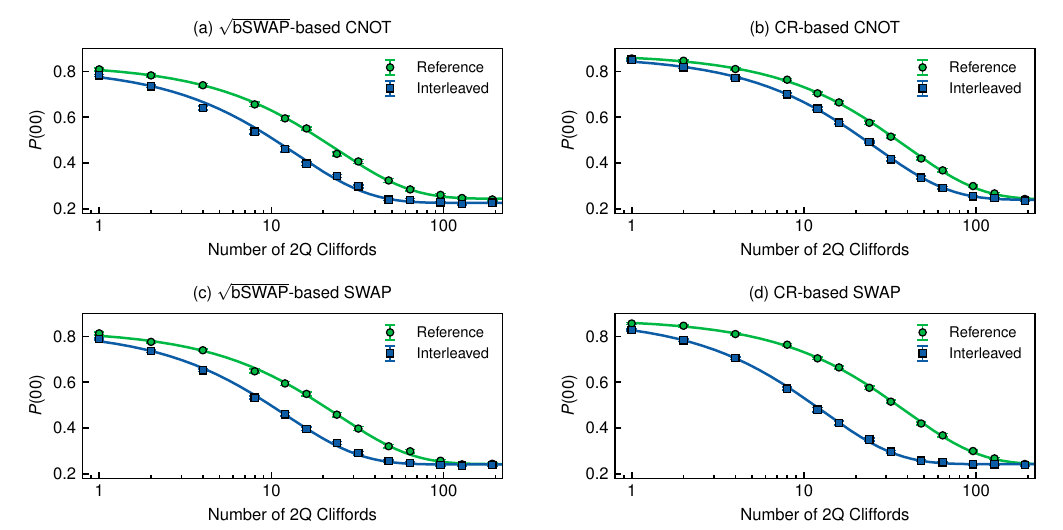}
\captionof{figure}{IRB of \xCNOT{} and \xSWAP{} implementations. (a)~$\xSQRTBSWAP$-based \xCNOT{}. (b)~CR-based \xCNOT{}. (c)~$\xSQRTBSWAP$-based \xSWAP{}. (d)~CR-based \xSWAP{}. Markers show mean ground-state populations over $32$ random sequences, with one-standard-error bars. Reference and interleaved circuits are shown in green and blue.}
\label{fig:operation_irb_comparison}
\end{center}
\twocolumngrid
\flushbottom

\end{document}